\documentclass[12pt,a4paper]{article}
\usepackage{graphicx}
\usepackage{amsmath}
\usepackage{cite}
\usepackage{epstopdf}

\usepackage{color}

\def\Comment#1{}
\newcommand{\kw}{\kappa_{W}}
\newcommand{\kV}{\kappa_{V}}
\newcommand{\bean}{\begin{eqnarray*}}
\newcommand{\eean}{\end{eqnarray*}}

\newcommand{\gapproxeq}{\lower
.7ex\hbox{$\;\stackrel{\textstyle >}{\sim}\;$}}
\newcommand{\lapproxeq}{\lower
.7ex\hbox{$\;\stackrel{\textstyle <}{\sim}\;$}}

\newcommand\lsim{\mathrel{\rlap{\lower4pt\hbox{\hskip1pt$\sim$}}
    \raise1pt\hbox{$<$}}}
\newcommand\gsim{\mathrel{\rlap{\lower4pt\hbox{\hskip1pt$\sim$}}
    \raise1pt\hbox{$>$}}}
\newcommand{\ba}{\begin{array}}
\newcommand{\ea}{\end{array}}
\newcommand{\nn}{\nonumber}

\newcommand{\be}{\begin{equation}}
\newcommand{\ee}{\end{equation}}
\newcommand{\bear}{\begin{eqnarray}}
\newcommand{\eear}{\end{eqnarray}}

\newcommand{\ket}{\,\rangle}
\newcommand{\bra}{\langle \,}
\newcommand{\eqn}[1]{(\ref{#1})}
\newcommand{\cO}{{\cal O}}
\newcommand{\bel}[1]{\be\label{#1}}
\newcommand{\mL}{\mathcal{L}}

\newcommand{\mF}{\mathcal{F}}

\newcommand{\mO}{\mathcal{O}}

\newcommand{\mS}{\mathcal{S}}

\def\bat{\begin{array}{cc}}
\newcommand{\toG}{\stackrel{G}{\,\longrightarrow\,}}
\newcommand{\Frac}[2]{\frac{\displaystyle #1}{\displaystyle #2}}
\newcommand{\Int}{\displaystyle{\int}}
\def\ie{{\it i.e.},\ }
\begin{document}
\thispagestyle{empty}
\begin{titlepage}
\begin{center}
\hfill FTUV/13$-$1011 \\
\hfill IFIC/13$-$57\\
\hfill FTUAM/13$-$23 \\
\hfill IFT-UAM/CSIC/13-093 \\

\vspace*{1.5cm}
\begin{Large}
{\bf
Oblique S and T Constraints on Electroweak \\[10pt]
Strongly-Coupled Models with a Light Higgs}
 \\[1.3cm]
\end{Large}

{ \sc A. Pich$^{1}$ }, {\sc I. Rosell$^{1,2}$} and { \sc J.J. Sanz-Cillero$^3$ }  \\[0.8cm]

{\it $^{1}$ Departament de F\'\i sica Te\`orica, IFIC, Universitat de Val\`encia --
CSIC\\
 Apt. Correus 22085, E-46071 Val\`encia, Spain }\\[0.3cm]

{\it $^{2}$Departamento de Ciencias F\'\i sicas, Matem\'aticas y de la Computaci\'on, ESET,\\
Universidad CEU Cardenal Herrera, 
E-46115 Alfara del Patriarca, Val\`encia, Spain }\\[0.3cm]

{\it $^3$
Departamento de F\'\i sica Te\'orica and Instituto de F\'\i sica Te\'orica, IFT-UAM/CSIC,
Universidad Aut\'onoma de Madrid, Cantoblanco, E-28049 Madrid, Spain
}\\[0.6cm]

\vspace*{0.9cm}
\begin{abstract}
Using a general effective Lagrangian implementing the chiral symmetry breaking $SU(2)_L\otimes SU(2)_R\to SU(2)_{L+R}$, we present a one-loop calculation of the oblique $S$ and $T$ parameters within electroweak strongly-coupled models with a light scalar. Imposing a proper ultraviolet behaviour, we determine $S$ and $T$ at next-to-leading order in terms of a few resonance parameters. The constraints from the global fit to electroweak precision data force the massive vector and axial-vector states to be heavy, with masses above the TeV scale, and suggest that  the $W^+W^-$ and $ZZ$ couplings of the Higgs-like scalar should be close to the Standard Model value. Our findings are  generic, since they only rely on soft requirements on the short-distance properties of the underlying strongly-coupled theory, which are widely satisfied in more specific scenarios.

\end{abstract}
\end{center}
\vfill
\eject
\end{titlepage}

\pagenumbering{arabic}

\parskip12pt plus 1pt minus 1pt
\topsep0pt plus 1pt
\setcounter{totalnumber}{12}


\section{Introduction}

The data accumulated so far \cite{Aad:2012tfa,Aad:2013wqa,Chatrchyan:2012ufa,Chatrchyan:2013lba,Aaltonen:2012qt} confirm the Higgs-like nature \cite{Higgs:1964pj,Englert:1964et,Guralnik:1964eu,Kibble:1967sv} of the new boson discovered at the LHC, with a spin and parity consistent with the Standard Model (SM) $0^+$ assignment \cite{Aad:2013xqa,Chatrchyan:2012jja,D0-6387}, and a mass $m_H=125.64\pm 0.35$~GeV \cite{Pich:2013xx}, in good agreement with the expectations from global fits to precision electroweak data \cite{Baak:2012kk,Eberhardt:2012gv}.
Although its properties are not well measured yet, the $H(126)$ boson
is a very compelling candidate to be the SM Higgs.
An obvious question to address is whether it corresponds to the unique Higgs boson incorporated in the SM, or it is just the first signal of a much richer scenario
of Electroweak Symmetry Breaking (EWSB).
Obvious possibilities are an extended scalar sector with additional fields or dynamical (non-perturbative) EWSB generated by some new underlying dynamics.

The SM implements the EWSB through a complex scalar doublet $\Phi(x)$ and
a potential $V(\Phi)$ with non-trivial minima, giving rise to
three Goldstone bosons which, in the unitary gauge, become the needed longitudinal polarizations of the gauge bosons. Since $\Phi(x)$ contains four real fields, one massive neutral scalar survives in the physical spectrum: the Higgs boson.
Although the Higgs is not needed for the EWSB, the scalar doublet structure provides a renormalizable model with good unitarity properties.
The scalar sector of the SM Lagrangian can be written in the form
\cite{Pich:1995bw,Pich:1998xt}
\bel{eq:l_sm}
\mL(\Phi)\, =\, {1\over 2}\, \langle\, \left(D^\mu\Sigma\right)^\dagger D_\mu\Sigma\,\rangle
- {\lambda\over 16} \left(\langle\,\Sigma^\dagger\Sigma\,\rangle
- v^2\right)^2 ,
\ee
where the $2\times 2$ matrix
\be
\Sigma \,\equiv\, \left( \Phi^c, \Phi\right) \, =\,
\left( \bat \Phi^{0*} & \Phi^+  \\ -\Phi^- &  \Phi^0 \ea\right)
\label{eq:sigma_matrix}
\ee
collects the scalar doublet and its charge-conjugate $\Phi^c = i \sigma_2\Phi^*$,
$\langle A\rangle$ stands for the trace of the $2\times 2$ matrix $A$, and
$D_\mu\Sigma \equiv \partial_\mu\Sigma
+ i g \,\frac{\vec{\sigma}}{2}\vec{W}_\mu \,\Sigma - i g' \,\Sigma \,\frac{\sigma_3}{2} B_\mu $
is the usual gauge-covariant derivative.
This expression makes manifest the existence of a global
$G\equiv SU(2)_L\otimes SU(2)_R$ symmetry,
\be
\Sigma \, \toG \,
g_L \,\Sigma\, g_R^\dagger , \qquad\qquad\qquad
g_{L,R}  \in SU(2)_{L,R} \, ,
\label{eq:sigma_transf}
\ee
which is broken by the vacuum to the diagonal $SU(2)_{L+R}$,
usually called custodial symmetry group \cite{Sikivie:1980hm}.
The SM promotes the $SU(2)_L$ to a local gauge symmetry, while only the
$U(1)_Y$ subgroup of $SU(2)_R$ is gauged; thus the $SU(2)_R$ symmetry is explicitly broken at $\cO(g')$ through the $U(1)_Y$ interaction in the covariant derivative.
Performing a polar decomposition,
\be
\Sigma(x) \, = \, {1\over\sqrt{2}}
\left[ v + H(x) \right] \, U(\varphi(x)) \, , \qquad\qquad\qquad
U(\varphi) \, =\,  \exp{\left\{ i \vec{\sigma} \,
\vec{\varphi} / v \right\} } \, ,
\label{eq:polar}
\ee
in terms of the Higgs field $H(x)$ and the Goldstones
$\vec{\varphi}(x)$,
one can rewrite $\mL(\Phi)$ in the form~\cite{Appelquist:1980vg,Longhitano:1980iz}:
\be
\mL(\Phi)\, =\, {v^2\over 4}\,
\langle\, D_\mu U^\dagger D^\mu U \,\rangle \, +\,
\cO\left( H \right) ,
\label{eq:sm_goldstones}
\ee
with
$D_\mu U \equiv \partial_\mu U
+ i g \,\frac{\vec{\sigma}}{2}\vec{W}_\mu \, U - i g'\, U \,\frac{\sigma_3}{2} B_\mu$.
In the unitary gauge $U=1$, this Lagrangian
reduces to the usual bilinear gauge-mass term, with
$Z^\mu \equiv \cos{\theta_W} W_3^\mu - \sin{\theta_W} B^\mu$,
$m_W = m_Z \,\cos{\theta_W} = v g/2$ and
$\tan{\theta_W} = g'/ g$.

Without the Higgs field, Eq.~\eqn{eq:sm_goldstones} is the generic lowest-order Goldstone Lagrangian associated with the symmetry breaking $SU(2)_L\otimes SU(2)_R\to SU(2)_{L+R}$. In Quantum Chromodynamics (QCD)
the same Lagrangian describes the dynamics of pions at $\cO (p^2)$
(two derivatives), with $v = f_\pi$ the pion decay constant \cite{Pich:1995bw}.
The successful electroweak precision tests of the SM~\cite{EW-rev-Pich}
have confirmed that this is also the right pattern of symmetry breaking associated with the electroweak Goldstone bosons,
with $v = \left(\sqrt{2} G_F\right)^{-1/2} = 246\:\mathrm{GeV}$.
The crucial question to be now investigated is whether the particular implementation of this symmetry breaking incorporated in the SM is the one chosen by Nature, with the $H(126)$ being the long-awaited Higgs boson.

The implications of the assumed Goldstone symmetry structure can be investigated,
independently of any particular implementation of the symmetry breaking,
applying the same momentum expansion techniques used in Chiral Perturbation Theory ($\chi$PT) to describe low-energy QCD \cite{Pich:1995bw,ChPT,ChPTp4,Ecker:1994gg}.
The electroweak Goldstone dynamics is then parameterized through an Effective Lagrangian which contains the SM gauge symmetry realized nonlinearly.
In the past
\cite{Appelquist:1980vg,Longhitano:1980iz,Dobado:1990zh,Espriu:1991vm,Dobado:1997jx},
only the known light degrees of freedom (leptons, quarks and gauge bosons) were included in the electroweak effective Lagrangian. The discovery of the $H(126)$ boson has triggered a renewed interest in this effective field theory approach, with a large number of works incorporating the Higgs-like boson as an explicit field in the effective low-energy Lagrangian \cite{Bagger:1993zf,BuchallaCK:2013,ContinoGGMS:2013,EWL_Higgs}.

We want to consider strongly-coupled models where the gauge symmetry is dynamically broken by means of some non-perturbative interaction.
Usually, theories of this kind do not contain any fundamental Higgs, bringing instead resonances of different types as happens in QCD
\cite{Chivukula:1998if,Pomarol:2012sb,Andersen:2011yj}.
For instance, Technicolour \cite{technicolor},
the most studied strongly-coupled model, introduces an asymptotically-free QCD replica at TeV energies which breaks the electroweak symmetry in the infrared,
in a similar way as chiral symmetry is broken in QCD. This gives rise to the appearance of a tower of heavy resonances in the scattering amplitudes.
Other models consider the possibility that the ultraviolet (UV) theory remains close to a strongly-interacting conformal fixed point over a wide range of energies (Walking Technicolour) \cite{walking}; recent work in this direction
incorporates conformal field theory techniques (Conformal Technicolour)
\cite{Luty:2004ye,Rattazzi:2008pe,new}.
Strongly-coupled models in warped \cite{Randall:1999ee} or deconstructed \cite{ArkaniHamed:2001ca} extra dimensions~\cite{Csaki:2003zu,SekharChivukula:2001hz,Agashe:2003zs,Agashe:2004rs,Contino:2006qr,composite,Contino:2010rs,Foadi:2003xa}
have been also investigated.

The $H(126)$ boson could indeed be a first experimental signal of a new strongly-interacting sector: the lightest state of a large variety of new resonances of different types. Among the many possibilities, the relatively light mass of the discovered Higgs candidate has boosted the interest on strongly-coupled scenarios with a composite pseudo-Goldstone Higgs boson \cite{Contino:2010rs,Espinosa:2010vn,Contino:2010mh},
where the Higgs mass is protected by an approximate global symmetry and is only generated via quantum effects \cite{Kaplan:1983fs}. A simple example is provided by the popular $\mathrm{SO(5)}/\mathrm{SO(4)}$ minimal composite Higgs model \cite{Agashe:2004rs,Contino:2006qr,composite,Contino:2010rs,Contino:2011,Marzocca:2012zn}.
One could also try to interpret the Higgs-like scalar as a dilaton, the pseudo-Goldstone boson associated with the spontaneous breaking of scale invariance at some scale $f_\varphi\gg v$ \cite{Goldberger:2007zk,Matsuzaki:2012xx,Bellazzini:2012vz,Chacko:2012vm,EP:2012}, and other plausible dynamical scenarios have been considered \cite{FoadiFS:2013}.

The dynamics of Goldstones and massive resonance states can be analyzed in a generic way by using an effective Lagrangian, based on symmetry considerations.
The theoretical framework is completely analogous to the Resonance Chiral Theory (R$\chi$T) description of QCD at GeV energies~\cite{RChT,Pich:2002xy}.
Using these techniques, we investigated in Ref.~\cite{S-Higgsless}  the oblique $S$ parameter~\cite{Peskin:92}, characterizing the electroweak boson self-energies, within Higgsless strongly-coupled models
at the next-to-leading order (NLO), {\it i.e.}, at one-loop.
We found that in most strongly-coupled scenarios of EWSB a high resonance mass scale is required, $M_V > 1.8$~TeV, to satisfy the stringent experimental limits.
The recent discovery of the $H(126)$ boson made mandatory to update the analysis, including the light-scalar contributions~\cite{PRL}.
In addition, we also presented the results of a corresponding one-loop calculation of the oblique $T$ parameter, which allowed us to perform a correlated analysis of both quantities~\cite{PRL}. Previous one-loop analyses within similar frameworks can be found in Refs.~\cite{Matsuzaki:2006wn,S-Isidori:08,S-Cata:10,other,S-Orgogozo:11,Orgogozo:2012}.

In this paper we describe in a deeper way the one-loop calculation of the
oblique $S$ and $T$ parameters, and analyze in detail the phenomenological implications for strongly-coupled models. We can profit from the experience acquired in low-energy QCD, where a thorough investigation of R$\chi$T at the one-loop level has been performed
in recent years~\cite{Cata:2001nz,L9a,L8,L10,L9,Natxo-thesis,RPP:05,RChT-EoM},
bringing an improved understanding of the resonance dynamics.
In particular, we make use of the procedure developed to compute the low-energy constants of $\chi$PT at NLO through a matching with R$\chi$T~\cite{Cata:2001nz,L9a,L8,L10,L9}.
The estimation of $S$ in strongly-coupled electroweak models is equivalent
to the calculation of $L_{10}$ in $\chi$PT~\cite{L10}, whereas the calculation of $T$ is similar to the determination of $f_{\pi^\pm}^2-f_{\pi^0}^2$ in $\chi$PT.
Previous one-loop estimates of $S$ and $T$ contained unphysical dependences on the UV cut-off, manifesting the need for local contributions to account for a proper UV completion. Our calculation avoids this problem through the implementation of short-distance conditions on the relevant Green functions, in order to satisfy the assumed UV behaviour of the strongly-coupled theory.
As shown in Refs.~\cite{L8,L10,L9}, the dispersive approach that we adopt
avoids all technicalities associated with the renormalization procedure,
allowing for a much more transparent understanding of the underlying physics.

The paper is organized as follows. The effective electroweak Lagrangian, including the singlet scalar and the lightest vector and axial-vector resonance multiplets,
is constructed in section~\ref{sec.lagrangian}.
In section~\ref{sec.observables}, we briefly review the definition of
the $S$ and $T$ parameters and the dispersive representation of $S$ advocated
by Peskin and Takeuchi \cite{Peskin:92}; we also explain there the dispersive relation we have used for the calculation of $T$.
Section~\ref{sec:LO} analyzes the lowest-order contributions to the oblique parameters and the implications of the short-distance constraints imposed by the UV behaviour of the underlying strongly-coupled theory.
The NLO computation of the parameter $S$ is presented in section~\ref{sec.NLO}, where we give a detailed description of the relevant spectral functions and implement a proper short-distance behaviour. Section~\ref{sec:T-NLO} describes the analogous calculation of the parameter $T$. The phenomenological implications are discussed in section~\ref{sec.pheno} and our conclusions are finally summarized in section~\ref{sec.conclusions}, where we also show briefly how they can be particularized to some popular models. Some technical aspects are given in the appendices.

\section{Electroweak effective theory}
\label{sec.lagrangian}

Let us consider a low-energy effective theory containing the SM gauge bosons coupled to the electroweak Goldstones, one scalar state $S_1$ with mass $m_{S_1} = 126$~GeV and the lightest vector and axial-vector resonance multiplets $V_{\mu\nu}$ and $A_{\mu\nu}$, which are expected to be the most relevant ones at low energies. We only assume the SM pattern of EWSB, {\it i.e.} the theory is symmetric under $G=SU(2)_L\otimes SU(2)_R$ and becomes spontaneously broken to the diagonal subgroup $H=SU(2)_{L+R}$. The scalar field $S_1$ is taken to be singlet under $SU(2)_{L+R}$, while $V_{\mu\nu}$ and $A_{\mu\nu}$ are triplets (singlet vector and axial-vector contributions are absent in our calculation, at the order we are working).
It is convenient to sort out the terms in the Lagrangian according to the
number of resonance fields:
\bear
\mL^{\rm EW} &=& \mL_G[W,B,\varphi]\, +\, \sum_R \mL_R[W,B,\varphi,R] \, +\
\sum_{R,R'} \mL_{RR'}[W,B,\varphi,R,R'] \, +\cdots
\eear
where $\mL_G$ contains terms without resonances (only Goldstones and gauge bosons), $\mL_R$ has one resonance of type $R$ ($R=S_1,V,A$), $\mL_{R,R'}$ two resonances, etc. In our calculation of the oblique parameters we only need terms with at most two resonance fields.

The Lagrangian could be further organized as an expansion in powers of derivatives (momenta) over the EWSB scale and one could write, in principle, operators with an arbitrary large number of derivatives. However, most higher-derivative operators are either redundant
(proportional to the equations of motion)~\cite{RChT-EoM} or do not contribute to the
vertices needed in our calculation. Moreover, operators with more than two derivatives
unavoidably lead to a highly-divergent behaviour of Green functions at high energies,
which is not allowed by the assumed short-distance constraints from the underlying
strongly-coupled theory, and must be discarded. Therefore, only operators with at most
two derivatives will be kept in the effective Lagrangian
(see appendix~\ref{app.RChT-EoM} for further details).

We will adopt a non-linear realization of the electroweak Goldstones~\cite{Appelquist:1980vg,Longhitano:1980iz}
and work out the most general operators in the Lagrangian allowed by the symmetry.
For the construction of the Lagrangian we will make use of the covariant tensors
\begin{eqnarray}
\hat{W}^{\mu\nu}\, =\, \partial^\mu \hat{W}^\nu  - \partial^\nu \hat{W}^\mu - i \, [\hat{W}^\mu,\hat{W}^\nu]\, ,
\qquad\qquad
\hat{B}^{\mu\nu}\, =\, \partial^\mu \hat{B}^\nu - \partial^\nu \hat{B}^\mu
- i \, [\hat{B}^\mu,\hat{B}^\nu]\, ,
\nn\\
u^\mu = i\, u\,  D^\mu U^\dagger\, u = -i\, u^\dagger  D^\mu U\, u^\dagger = u^{\mu\dagger}\, ,
\qquad\qquad\;
D^\mu U = \partial^\mu U - i \, \hat{W}^\mu U+ i \,U\, \hat{B}^\mu\, .\;\;
\end{eqnarray}
The Goldstone bosons are parameterized through
$U=u^2=\exp{\left\{ i \vec{\sigma} \vec{\varphi} / v \right\} }$,
where $u(\varphi)$ is an element of the coset $G/H$.
Under a transformation\ $g\equiv (g_L,g_R)\in G$,\footnote{
For a given choice of coset representative $\bar\xi(\varphi)\equiv \left(\xi_L(\varphi),\xi_R(\varphi)\right)\in G$, the change of the
Goldstone coordinates under a chiral transformation takes the form
$$ \xi_L(\varphi) \to g_L\, \xi_L(\varphi)\, h^\dagger(\varphi,g)\, ,
\qquad\qquad\qquad
\xi_R(\varphi) \to g_R\, \xi_R(\varphi)\, h^\dagger(\varphi,g)\, . $$
The same compensating transformation $h(\varphi,g)$ occurs in both chiral sectors because
they are related by a discrete parity transformation $L\leftrightarrow R$ which leaves
$H$ ($L+R$) invariant. $U(\varphi)\equiv \xi_L(\varphi)\xi_R^\dagger(\varphi)$ transforms
as $g_L\, U(\varphi)\,g_R^\dagger$.
We take a canonical choice of coset
representative such that $\xi_L(\varphi) = \xi_R^\dagger(\varphi)\equiv u(\varphi)$.
}
%
\be
u(\varphi)\quad \longrightarrow\quad g_L \, u(\varphi)\, h^\dagger(\varphi,g)\,=\,  h(\varphi,g) \, u(\varphi)\,  g_R^\dagger\, ,
\ee
with $h\equiv h(\varphi,g)\in H$ a compensating transformation to
preserve the coset representative \cite{Coleman:1969sm}.
Requiring the $SU(2)$ matrices $\hat{W}^\mu$ and $\hat{B}^\mu$ to transform as
\bel{eq:FakeTransform}
\hat{W}^\mu\to g_L\, \hat{W}^\mu g_L^\dagger + i\, g_L\, \partial^\mu g_L^\dagger\, ,
\qquad\qquad
\hat{B}^\mu\to g_R\, \hat{B}^\mu g_R^\dagger + i\, g_R\, \partial^\mu g_R^\dagger\, ,
\ee
the effective Lagrangian is invariant under local $SU(2)_L\otimes SU(2)_R$
transformations. The identification
\bel{eq:SMgauge}
\hat{W}^\mu \, =\, -g\;\frac{\vec{\sigma}}{2}\, \vec{W}^\mu \, ,
\qquad\qquad\qquad
\hat{B}^\mu\, =\, -g'\;\frac{\sigma_3}{2}\, B^\mu\, ,
\ee
breaks explicitly the $SU(2)_R$ symmetry group, in exactly the same way as the
SM does, preserving the $SU(2)_L\otimes U(1)_Y$ gauge symmetry.
Taking functional derivatives with respect to the formal left and right sources $\hat{W}^\mu$ and $\hat{B}^\mu$, one can also study the corresponding currents (and current Green functions).

The inner nature of the EWSB is left unspecified. Instead of the SM Higgs,
we assume that the strongly-coupled underlying dynamics gives rise to massive resonance multiplets transforming as triplets ($R\equiv \frac{\vec{\sigma}}{\sqrt{2}}\, \vec{R}$) or singlets ($R_1$) under $H$:
\be
R\quad \longrightarrow\quad h(\varphi,g) \: R\: h^\dagger(\varphi,g)\, ,
\qquad\qquad\qquad
R_1\quad \longrightarrow\quad R_1\, .
\ee
In order to build invariant operators under the assumed symmetry group,
it is useful to introduce \cite{RChT} the covariant derivative
\be
\nabla^\mu R \, =\, \partial^\mu R +  \left[ \Gamma^\mu , R\right]\, ,
\qquad\qquad
\Gamma^\mu \,=\, \Frac{1}{2}  \left\{u \left( \partial^\mu
- i \hat{B}^\mu \right) u^\dagger +u^\dagger \left(\partial^\mu - i \hat{W}^\mu \right) u \right\}
\, ,
\ee
and
\be
h^{\mu\nu}\, =\, \nabla^\mu u^\nu + \nabla^\nu u^\mu \, ,
\qquad\qquad
f^{\mu\nu}_{\pm} \, =\,
u^\dagger \,\hat{W}^{\mu\nu} u \pm  u \,\hat{B}^{\mu\nu} u^\dagger \, ,
\ee
which transform as triplets under $H$:
\be
\left\{ \nabla^\mu R\, ,\, h^{\mu\nu}\, ,\, f^{\mu\nu}_{\pm}\, ,\, u^\mu\right\}
\quad \longrightarrow\quad h \; \left\{ \nabla^\mu R\, ,\, h^{\mu\nu}\, ,\, f^{\mu\nu}_{\pm}\, ,\, u^\mu\right\}\; h^\dagger\, .
\ee

In our general $SU(2)_L \otimes SU(2)_R$ framework the terms with no resonance fields are
\begin{equation}
\mathcal{L}_G   \;=\;
-\frac{1}{2g^2}\,\bra \hat{W}_{\mu\nu} \hat{W}^{\mu\nu} \ket
\, -\,\frac{1}{2g'^{\, 2}}\,\bra  \hat{B}_{\mu\nu} \hat{B}^{\mu\nu} \ket
\, +\, \frac{v^2}{4}\, \bra u_\mu u^\mu \ket \, ,
\label{eq.L_G}
\end{equation}
which provide the usual Yang-Mills action and the last term in the equation is  the Goldstone Lagrangian in Eq.~(\ref{eq:sm_goldstones}).
The Lagrangians with resonance fields can be directly taken from Refs.~\cite{RChT}, with minimal notational changes. We just consider a
singlet scalar $S_1$ and the lowest-mass vector ($V^{\mu\nu}$)
and axial-vector ($A^{\mu\nu}$) triplet multiplets,
which can induce sizeable corrections to the gauge-boson self-energies.
We use the antisymmetric tensor formalism\footnote{
In addition to provide the same type of description for vector and axial-vector states, this formalism avoids the mixing of the axial resonances with the Goldstones and its softer momentum dependence allows us to recover in a simpler way the right UV behaviour.
Alternative realizations with spin--1 resonances in the Proca formalism can be found in~\cite{Alboteanu:2008my}, together with spin--2 fields and
higher dimension multiplets of the electroweak symmetry group.}
to describe these spin--1 fields \cite{ChPTp4,RChT}
and assume that the strong dynamics
preserves parity ($L\leftrightarrow R$) and charge conjugation. For our calculation, we will need the following operators with one resonance field,
\bear\label{eq:L_R}
\mL_{S_1} + \mathcal{L}_A+ \mL_V &=&
\Frac{1}{2}\, \kw\, v \, S_1 \,\bra  u_\mu u^\mu \ket\,
 + \frac{F_A}{2\sqrt{2}}\, \bra A_{\mu\nu} f^{\mu\nu}_- \ket
\nonumber \\ &&\mbox{}
+ \frac{F_V}{2\sqrt{2}}\, \bra V_{\mu\nu} f^{\mu\nu}_+ \ket
+ \frac{i\, G_V}{2\sqrt{2}}\, \bra  V_{\mu\nu} [u^\mu, u^\nu] \ket
\, ,
\eear
and only one operator with two resonances, involving the singlet scalar boson and the axial multiplet:
\be\label{eq:L_RRp}
\mL_{S_1 A} \; =\;
\sqrt{2}\, \lambda_1^{SA}\,  \partial_\mu S_1 \, \bra A^{\mu \nu} u_\nu \ket\, .
\ee

The term proportional to $\kw$ in Eq.~(\ref{eq:L_R}) contains the coupling of the scalar $S_1$ resonance to two gauge bosons. Since it respects custodial symmetry, $\kw$ parametrizes both the $S_1W^+W^-$ and $S_1ZZ$ couplings.\footnote{
The coupling $\kw$ was denoted as $\omega$ in Refs~\cite{S-Higgsless,PRL}. In other  references it is also called $a$ or $\kV$.}
For later convenience we will take the sign convention $\kw\geq 0$; there is no loss of generality, if one does not demand other scalar couplings to be positive {\it a priori},
as one is always allowed to perform a flip of sign in this particular coupling
through a global field redefinition $S_1\to - S_1$.
For $\kw=1$ one recovers the $S_1\to\varphi\varphi$ vertex of the SM.

Collecting all pieces, the effective Lagrangian we are going to use reads
\be
\mathcal{L} \; =\; \mathcal{L}_{G}
\,+\, \mathcal{L}_{\mathrm{GF}}
\,+ \, \mL_{S_1}+\mathcal{L}_{V}+\mathcal{L}_{A}
\, +\, \mL_{S_1 A} \,+\,\mathcal{L}_{S_1S_1}^{\mathrm{kin}}
\, +\, \mathcal{L}_{VV}^{\mathrm{kin}} \,+\, \mathcal{L}_{AA}^{\mathrm{kin}}
\, ,
\ee
with
\be
\mathcal{L}_{\mathrm{GF}}\, =\, -\frac{1}{2\xi}\, (\partial^\mu \vec{W}_\mu)^2
\ee
the gauge-fixing term. The calculation of the oblique $S$ and $T$ parameters will be performed in the Landau gauge $\xi \to  0$, so that the gauge boson propagators are transverse. This eliminates any possible mixing of the Goldstones and the gauge bosons, which can only occur through the longitudinal parts of the $W^\pm$ and $Z$ propagators.

\section{Oblique parameters}
\label{sec.observables}

The $Z$ and $W^\pm$ self-energies are modified by the presence of massive resonance states coupled to the gauge bosons. The deviations with respect to the SM predictions are characterized by the so-called oblique parameters
\cite{Peskin:92,Kennedy:1990ib,S-def_Barbieri,S-def_Pomarol}.
The leading effects on precision electroweak measurements are described in terms of three parameters $S$, $T$ and $U$ (or equivalently $\varepsilon_1$, $\varepsilon_2$ and $\varepsilon_3$), but most simple types of new physics give $U=0$, which we will not discuss any further.
$S$ measures the difference between the off-diagonal $W^3 B$ correlator and its SM value, while $T$ parametrizes the breaking of custodial symmetry.
Their precise definitions involve the quantities
\begin{equation}
e_3\,=\, \Frac{g}{g'}  \; \widetilde{\Pi}_{30}(0) \, ,
\qquad\qquad
e_1\,=\,
\frac{ \Pi_{33}(0) - \Pi_{WW} (0)}{M_W^2}\,,
\label{eq.S-def}
\end{equation}
where the tree-level Goldstone contribution has been removed from $\Pi_{30}(q^2)$
in the form \cite{Peskin:92}:
\begin{equation}
\Pi_{30}(q^2)\,=\,q^2\, \widetilde\Pi_{30}(q^2)\,+\,\frac{g^2 \tan{\theta_W}}{4}\, v^2 \,  .
\label{eq.PiTilde}
\end{equation}
The $S$ and $T$ parameters are given by the deviation
with respect to the SM contributions  $e_3^{\rm SM}$ and $e_1^{\rm SM}$,
respectively:
\begin{equation}
S\,=\,  \Frac{16\pi}{g^2}\;\big(e_3 - e_3^{\rm SM}\big)\, ,
\qquad \qquad
T\,=\, \Frac{4\pi  }{g^2   \sin^2{\theta_W}}\; \big(e_1-e_1^{\rm SM}\big)  \,.
\end{equation}

In order to define the SM contribution, and therefore $S$ and $T$,
one needs a reference value for the SM Higgs mass. Taking $m_H = 126$~GeV, the global fit to precision electroweak data \cite{Baak:2012kk} gives the results
\be\label{eq:S_T_ewfit}
S\; =\; 0.03\pm 0.10\, ,
\qquad\qquad\qquad
T\; =\; 0.05\pm 0.12\, ,
\ee
with a correlation coefficient of 0.891.

A useful dispersive representation for the $S$ parameter was  introduced by
Peskin and Takeuchi~\cite{Peskin:92}:
\bear
S &=& \Frac{16\pi}{g^2 \tan\theta_W}\; \Int_0^\infty \Frac{{\rm dt}}{t}\;
[\rho_S(t)\, -\,\rho_S(t)^{\rm SM} ]\, , \label{Peskin-Takeuchi}
\eear
with the spectral function
\bear
\rho_S(t) &=&\Frac{1}{\pi}\,\mbox{Im}\widetilde{\Pi}_{30}(t)\, .
\eear
In the SM one has at one-loop (we will work at lowest order in $g$ and $g'$)
\bear
\rho_S(s)^{\rm SM} &=&  \frac{g^2 \tan\theta_W}{192\pi^2}\, \left[ \theta(s) \,
-\,  \left(1-\frac{m_H^2}{s} \right)^3 \theta \left(s-m_H^2 \right) \right] \, .
\eear

The convergence of this unsubtracted dispersion relation requires a vanishing
spectral function at short distances.
In the SM, $\mathrm{Im}\widetilde{\Pi}_{30}(s)$ vanishes at $s\to\infty$
due to the interplay of the two-Goldstone and  the Goldstone--Higgs contributions. We will see later that this UV convergence is realized in a different way in electroweak strongly-coupled theories.
The dispersion relation allows us to avoid the computation of non-absorptive loop diagrams, which may be out of the reach of our effective Lagrangian description, as one should add many more terms allowed by symmetry to pin them down properly.
Furthermore, the requirement that the spectral function must vanish at high energies and the integral must be convergent removes from the picture any undesired UV cut-off. Thus, the determination of $S$ only depends
on the physical scales present in the problem.

The $1/t$  weight enhances the contribution from the lightest thresholds and suppresses channels with heavy states~\cite{L10}. Thus, we will focus our attention here on the lightest one and two-particle cuts: $\varphi$, $V$, $A$, $\varphi\varphi$ and $S_1\varphi$.
Since the leading-order (LO) determination of $S$ already implies
that the vector and axial-vector masses must be above the TeV scale
(see section~\ref{sec:LO}), two-particle cuts with $V$ and $A$ resonances are very
suppressed; their effect was estimated in Ref.~\cite{S-Higgsless} and found to be small.
For the same reason, we neglect contributions from possible fermionic resonances,
present in many beyond-SM models, which could only appear at the loop level and, {\it a priori}, are expected to be suppressed by their heavier thresholds. These kind of contributions have been analyzed 
in previous works~\cite{Agashe:2004rs,Anastasiou:2009rv,Grojean:2013qca,Barbieri:2007bh,Barbieri:2008zt,Golden:1990ig,Azatov:2013ura}, where fermionic loops 
were estimated, finding sizable corrections for some types of models.  
Although the fermion couplings and masses  have been thoroughly 
studied~\cite{Anastasiou:2009rv,Grojean:2013qca},   the loop estimates  usually rely 
on dimensional analysis  and/or  the use of UV cut-offs~\cite{Agashe:2004rs,Barbieri:2007bh,Barbieri:2008zt,Golden:1990ig,Azatov:2013ura}.   
A full EFT computation accounting for counter-terms and systematic  renormalization 
should be the aim of future analyses along this line 
and is out of the scope of this article. 

For the computation of $T$, we will use the Ward-Takahashi identity worked out
in Ref.~\cite{Barbieri:1992dq}. In the Landau gauge, instead of studying the
more cumbersome correlators $\Pi_{33}$ and $\Pi_{WW}$,
one simply needs to compute the self-energies of the
electroweak Goldstones~\cite{Barbieri:1992dq}:
\bear
e_1 &=& \Frac{Z^{(+)}}{Z^{(0)}} \, -\, 1\, \,\, \simeq  \,\,\,
\Sigma'(0)^{(0)}\, -\, \Sigma'(0)^{(+)}\, .
\label{eq.Ward-id}
\eear
The constants  $Z^{(+)}$ and $Z^{(0)}$ are the wave-function renormalizations
for the charged and neutral Goldstones, respectively. More precisely, they are provided by the derivative of the Goldstone self-energies at zero momentum:
$Z^{(i)}=1-\Sigma'(0)^{(i)}$, with $\Sigma'(t)\equiv\mathrm{d}\Sigma(t)/\mathrm{d}t$.
This leads to the second identity in~(\ref{eq.Ward-id}), which holds as far as the calculation remains at the NLO.

We will present later the one-loop contributions to $T$ from the lightest two-particle cuts: $\varphi B$ and $S_1 B$. Our analysis of these contributions shows that, once proper short-distance conditions have been imposed, the spectral function of the Goldstone self-energy difference,
\be
\rho_T(t)\,=\, \frac{1}{\pi}\mbox{Im}[\Sigma(t)^{(0)}-\Sigma(t)^{(+)}]\, ,
\ee
vanishes at high energies. Hence, one is allowed to recover the low-energy value of the self-energy difference and the $T$ parameter by means of the converging dispersion relation
\bear
T &=& \Frac{4\pi}{ g'^{\, 2} \cos^2\theta_W} \; \Int_0^\infty \Frac{{\rm dt}}{t^2}
\; [\rho_T(t)\, -\, \rho_T(t)^{\rm SM}]\, ,
\label{eq.T-disp-rel}
\eear
where the SM one-loop spectral function reads
\begin{eqnarray}
\rho_T(s)|_{SM} &=& \Frac{3g'^{\, 2}s}{64\pi^2}\;  \bigg[-\theta(s)+\left(1-\frac{m_H^4}{s^2} \right)\theta (s-m_H^2)\bigg] \,. \label{T-dispersive}
\end{eqnarray}
We stress that this property has only been checked explicitly for the leading $\varphi B$ and $S_1 B$ contributions.

\section{LO calculation and short-distance constraints}
\label{sec:LO}

\begin{figure}
\begin{center}
\includegraphics[scale=0.8]{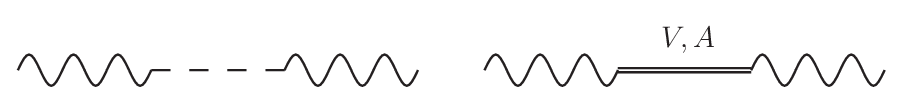}
\caption{\small{LO contributions to
$\Pi_{30}(s)$. A dashed line stands for a Goldstone boson, a double line indicates a resonance field and a curved line represents a gauge boson.}} \label{LO_graphs}
\end{center}
\end{figure}

The $T$ parameter vanishes at lowest order, because the tree-level Goldstone self-energies are identically zero and the corresponding wave-function renormalizations are $Z^{(k)}=1$:
\be
T_{\rm LO}\,=\, 0\, .
\ee

The oblique $S$ parameter receives tree-level contributions from vector and axial-vector exchanges.
The tree-level contributions to the gauge-boson vacuum polarization $\Pi_{30}(s)$
are shown in Figure~\ref{LO_graphs} and lead to the well-known LO result
\cite{Peskin:92} 
\begin{equation}
\left. \Pi_{30}(s) \right|_{\mathrm{LO}}\; =\;
\frac{g^2  \tan{\theta_W} }{4}\; s\;  \left(\frac{v^2}{s}+  \frac{F_V^2}{M_V^2-s}
- \frac{F_A^2}{M_A^2-s} \right)\, .
\label{eq.TL}
\end{equation}
The first term contains the Goldstone pole, which determines $\Pi_{30}(0)$.
This constant piece (also present in the SM) has been subtracted in
the definition of $\widetilde\Pi_{30}(s)$ in
Eqs.~(\ref{eq.S-def}) and (\ref{eq.PiTilde})
and does not play any role in the $S$ parameter:
\be\label{eq.S_LO}
S_{\mathrm{LO}}
\,=\, 4\pi\, \left( \frac{F_V^2}{M_V^2} - \frac{F_A^2}{M_A^2} \right)\,.
\ee
The result can be trivially generalized to incorporate the exchange
of several vector and axial-vector resonance multiplets~\cite{Knecht:1997ts}.

\subsection{Weinberg sum rules}

Since we are assuming that weak isospin and parity are good symmetries of the strong
dynamics, the correlator $\Pi_{30}(s)$ can be written in terms of the vector ($R+L$) and axial-vector ($R-L$) two-point functions as \cite{Peskin:92}
\be
\Pi_{30}(s)\, =\, \frac{g^2 \tan{\theta_W}}{4}\; s\;
\left[ \Pi_{VV}(s) - \Pi_{AA}(s)\right]\, .
\ee
The short-distance behaviour of this difference can be analyzed through the
Operator Product Expansion (OPE) of the right and left currents.
Owing to the chiral symmetry of the underlying theory, the only non-zero contributions
involve order parameters of the EWSB, \ie operators invariant under $H$ but not under $G$.
This guarantees the convergence of the dispersion relation
\eqn{Peskin-Takeuchi}  
because the unit operator is obviously symmetric.
In asymptotically-free gauge theories the difference $\Pi_{VV}(s)-\Pi_{AA}(s)$ vanishes at $s\to\infty$ as $1/s^3$ \cite{Bernard:1975cd}. This implies two super-convergent sum rules,
known as the first and second Weinberg sum rules (WSRs)~\cite{WSR}:
\begin{eqnarray}
\frac{1}{\pi}\,\int_0^\infty\;\mathrm{d}t\;
\left[\mathrm{Im}\Pi_{VV}(t)-\mathrm{Im}\Pi_{AA}(t)\right] &=& v^2\, ,
\\[10pt]
\frac{1}{\pi}\,\int_0^\infty\; \mathrm{d}t\; t\;
\left[\mathrm{Im}\Pi_{VV}(t)-\mathrm{Im}\Pi_{AA}(t)\right] &=& 0\, .
\end{eqnarray}
It is likely that the first of these sum rules is also true in gauge theories
with non-trivial UV fixed points.\footnote{
The specific condition required is that the OPE of $\Pi_{VV}(s)-\Pi_{AA}(s)$
does not contain operators with physical scaling dimension as low as 4
(for the second sum rule) or 2 (for the first)  \cite{Peskin:92}.}
However, the second WSR cannot be used in Conformal Technicolour models \cite{S-Orgogozo:11}
and its validity is questionable in most Walking Technicolour scenarios~\cite{Appelquist:1998xf}.

{}From the short-distance expansion of Eq.~(\ref{eq.TL}), one easily obtains the
implications of the WSRs at LO. The first WSR imposes the relation
\begin{equation}
 F_{V}^2 \,-\, F_{A}^2 \, =\, v^2 \, ,
\label{eq:1stWSR-LO}
\end{equation}
while requiring $ \Pi_{30}(s)$ to vanish as $1/s^2$  at short distances
(second WSR) leads to
\begin{equation}
F_{V}^2 \, M_{V}^2\, -\, F_{A}^2 \, M_{A}^2  \,=\, 0 \,  .
\label{eq:2ndWSR-LO}
\end{equation}
Therefore, if both WSRs are valid, $M_A > M_V$ and
the vector and axial-vector couplings are determined at LO in terms of the resonance masses:
\begin{equation}
\label{eq:FV_FA}
F_V^2\, =  \, v^2\, \Frac{M_A^2}{M_A^2-M_V^2} \, ,
\qquad \qquad \qquad F_A^2\, =\,v^2\, \frac{M_V^2}{M_A^2-M_V^2}\, .
\end{equation}

\subsection{Phenomenological implications}

Let us now analyze the impact of the previous short-distance constraints on the
LO prediction for the $S$ parameter in Eq.~\eqn{eq.S_LO}:

\begin{enumerate}
\item If one assumes the validity of the two WSRs, $F_V$ and $F_A$ take the
values in Eq.~(\ref{eq:FV_FA}), and $S_{\mathrm{LO}}$ becomes~\cite{Peskin:92}
\begin{equation}
S_{\mathrm{LO}}\; =\; \frac{4\pi v^2}{M_V^2}\,  \left( 1 + \frac{M_V^2}{M_A^2} \right) \, .
\label{eq.LO-S+2WSR}
\end{equation}
Since the WSRs imply $M_A>M_V$, the prediction
turns out to be bounded by~\cite{S-Higgsless}
\begin{equation}
\frac{4\pi v^2}{M_V^2}\;\,\mathrm{Max}\!\left(1,\frac{2 M_V^2}{M_A^2}\right)
\; < \; S_{\rm   LO} \; < \;   \frac{8 \pi v^2}{M_V^2} \, . \label{SLOtwoWSR}
\end{equation}

\item If only the first WSR is considered, and assuming  $M_A>M_V$, one obtains the lower bound~\cite{S-Higgsless}
\begin{equation}
S_{\mathrm{LO}}
\; =\; 4\pi \left\{ \frac{v^2}{M_V^2} + F_A^2 \left( \frac{1}{M_V^2} - \frac{1}{M_A^2} \right)
\right\}\; >\; \frac{4\pi v^2}{M_V^2}  \; >\; \frac{4\pi v^2}{M_A^2}    \, .
\label{eq.LO-S+1WSR}
\end{equation}
Thus, $S_{\mathrm{LO}}$ is predicted to be positive, provided $M_A>M_V$.

The possibility of an inverted mass ordering of the vector and axial-vector resonances in vector-like $SU(N)$ gauge theories, close to a conformal transition region, was considered in~\cite{Appelquist:1998xf}.
Composite models with one vector and two axial-vector resonances
also find allowed configurations with a similar inverted hierarchy~\cite{Marzocca:2012zn}.
If $M_V>M_A$, instead of a lower bound, the first identity in Eq.~\eqn{eq.LO-S+1WSR} implies the upper bound:
$S_{\mathrm{LO}} < 4\pi v^2/M_V^2 < 4\pi v^2/M_A^2 $.
In the degenerate mass limit $M_V=M_A$ all the inequalities would become identities. Thus, if the splitting of the vector and axial-vector resonances is small, the prediction of
$S_{\rm LO}$ would be close to the upper
bound and the main conclusion of this section would be stable.
\end{enumerate}

\begin{figure}[t]
\begin{center}
\includegraphics[scale=1,width=7.75cm]{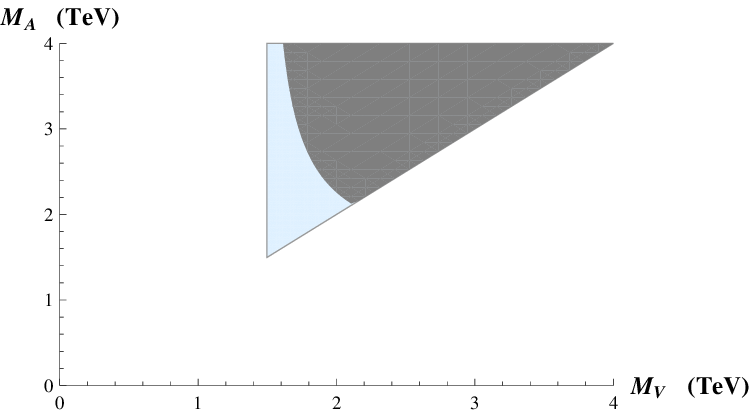}
\hskip .5cm
\includegraphics[scale=1,width=7.75cm]{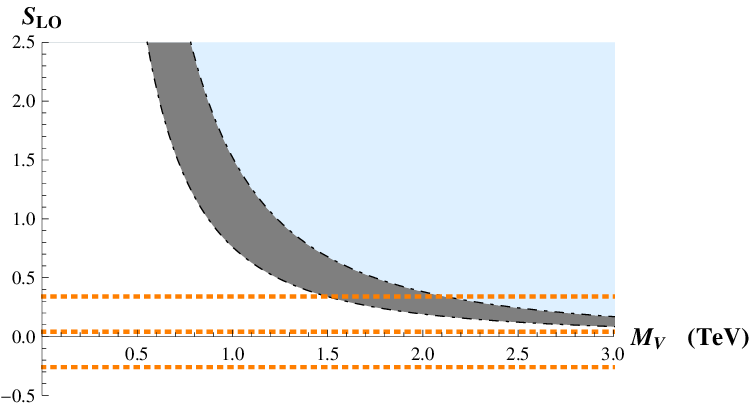}
\caption{\small
Regions for $M_V$ and $M_A$ where $S_{\mathrm{LO}}$ is compatible with the
data at the $3\sigma$ level (left) and LO predictions for $S$ (right).
The dark gray regions assume the two WSRs, while the light-blue areas
only require the first WSR and $M_A>M_V$.
The horizontal dotted lines in the right panel correspond to the experimentally allowed region at $3\sigma$.}
\label{fig.allowed-LO}
\end{center}
\end{figure}

The resonance masses need to be heavy enough to comply with the stringent experimental limits on $S$, in Eq.~\eqn{eq:S_T_ewfit}.
Figure~\ref{fig.allowed-LO} shows the ranges of resonance masses, $M_V$ and $M_A$, which are compatible with the experimental data at the $3\sigma$ level.
The dark gray region assumes the two WSRs, while the allowed range gets enlarged to the light-blue region if the second WSR is relaxed, and one only assumes the first WSR and $M_A> M_V$.
Even with the softer requirements, the experimental data implies $M_V>1.5$~TeV (2.3~TeV) at the 3$\sigma$ (1$\sigma$) level.
The right panel compares the corresponding LO predictions for $S$ with the experimentally allowed region at 3$\sigma$.

\section{NLO calculation of $S$}
\label{sec.NLO}

The experimental constraints on the $S$ parameter refer to a given reference value of the Higgs mass. However, the SM Higgs contribution only appears at the one-loop level. Thus, there is a scale ambiguity when comparing the LO theoretical result with the experimental constraint. This is similar to what happens in QCD with the tree-level estimate of the analogous parameter $L_{10}$, which does not capture its renormalization-scale dependence. In both cases, a one-loop calculation is needed to fix the ambiguity~\cite{ChPTp4,Dobado:1990zh}.

The NLO contribution is most efficiently obtained through a dispersive calculation. The essential condition needed to properly define the Peskin-Takeuchi representation in Eq.~(\ref{Peskin-Takeuchi}) is a vanishing spectral function
$\mathrm{Im}\widetilde{\Pi}_{30}(s)$ at $s\to\infty$; \ie
the correlator $\Pi_{30}(s)$ should behave at most as a constant at short distances. We have already seen in the previous section that this condition is indeed fulfilled in any strongly-coupled theory satisfying our assumed pattern of EWSB. This allows us to reconstruct the correlator from the spectral function:
\be\label{eq.dispersivePi}
\Pi_{30}(s)\,= \, \Pi_{30}(0)\, +\, \Frac{s}{\pi}\,\int_0^\infty \, \Frac{\mathrm{d}t}{t\,(t-s)}\;
\mbox{Im}\Pi_{30}(t)\, .
\ee
The subtraction constant $\Pi_{30}(0)$ is fixed by the Goldstone-pole contribution in Eq.~\eqn{eq.PiTilde}.
Some care has to be taken with the simultaneous presence of resonance poles and two-particle cuts. For simplicity, we omit here all technical aspects concerning the dispersive integral and the integration circuit. A more precise discussion is given in appendix~A of Ref.~\cite{S-Higgsless}.

\begin{figure}
\begin{center}
\includegraphics[scale=0.5]{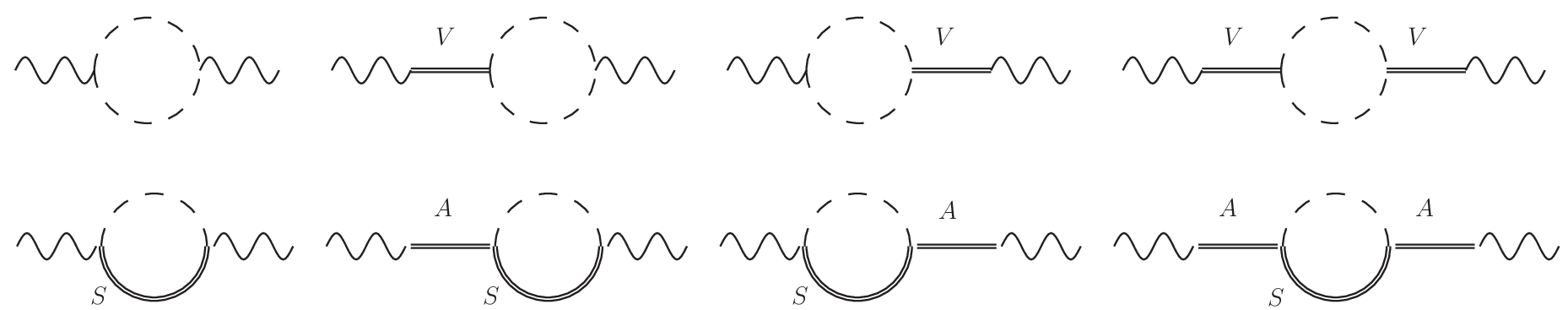}
\caption{\small{NLO contributions to $\mbox{Im} \Pi_{30}(s)$.
A dashed line stands for a Goldstone boson,
a double line indicates a resonance field ($V,\, A,\, S_1$)
and a curved line represents a gauge boson.}}
\label{fig.S-NLO_graphs}
\end{center}
\end{figure}

Figure~\ref{fig.S-NLO_graphs} shows the one-loop contributions to $\Pi_{30}(s)$ generating absorptive parts. We have considered two-particle cuts with two Goldstones or one Goldstone plus one scalar resonance. The two Goldstone contribution is also present in the SM and, therefore, its leading component cancels out from the $S$ parameter; this guarantees the good infrared behaviour of the representation (\ref{Peskin-Takeuchi}). We neglect the absorptive contributions from higher-mass two-particle cuts, which are kinematically suppressed by their much heavier thresholds. The $V\varphi$ and $A\varphi$ contributions were already analyzed in the Higgsless scenario and found to be small~\cite{S-Higgsless}.

Using the once-subtracted dispersion relation for $\Pi_{30}(s)$,
the total NLO result, including the tree-level exchanges, can be written in
the form~\cite{L10,Natxo-thesis,S-Higgsless}
\be
\left. \Pi_{30}(s) \right|_{\mathrm{NLO}}  \; =\;
\frac{g^2\tan{\theta_W} }{4}\;  s \;  \left(\frac{v^2}{s}+  \frac{F_{V}^{r\,2}}{M_{V}^{r\,2}-s}
- \frac{F_{A}^{r\,2}}{M_{A}^{r\,2}-s} \; +\; \overline{\Pi}(s)\right)\, ,
\label{eq.T-NLO}
\ee
where $F_R^r$ and $M_R^r$ are renormalized couplings which properly
define the resonance poles at the one-loop level. The one-loop contribution from the two-particle cuts is contained in $\overline{\Pi}(s)$. The precise definition of $\overline{\Pi}(s)$ is given in appendix~A of Ref.~\cite{S-Higgsless}.
At NLO the predicted $S$ parameter takes the form
\be\label{eq.Sbar}
S \; =\;
4\pi \left( \frac{F_{V}^{r\,2}}{M_{V}^{r\,2}} - \frac{F_{A}^{r\,2}}{M_{A}^{r\,2}} \right)
\; +\; \overline{S} \, ,
\ee
with $\overline{S}=4\pi\, \overline{\Pi}(0)$.

\subsection{Spectral functions}

The two-Goldstone and $S_1\varphi^0$ contributions to the spectral function are given by
\bear
\rho_S(s)|_{\varphi\varphi}  &=& \theta(s)\;
\Frac{g^2\tan\theta_W}{192\pi^2}\; |\mF^v_{\varphi\varphi}(s)|^2 \, ,
\\
\rho_S(s)|_{S_1\varphi}  &=&\mbox{} -\theta(s-m_{S_1}^2)\;
\Frac{g^2\tan\theta_W}{192\pi^2}\; |\mF^a_{S_1\varphi}(s)|^2
\;\left(1-\Frac{m_{S_1}^2}{s}\right)^3\, ,
\eear
where $\mF^v_{\varphi\varphi}(s)$ and $\mF^a_{S_1\varphi}(s)$ are the so-called $\varphi\varphi$ vector and $S_1\varphi$ axial form factors, respectively, defined by the corresponding matrix elements of the vector and axial-vector currents. At LO, they get a direct constant contribution plus a resonance-exchange term proportional to $\sigma_V\equiv F_V G_V/v^2$~\cite{RChT} and $\sigma_A\equiv F_A \lambda_1^{\mathrm{SA}}/(\kw v)$~\cite{L10,Natxo-thesis},\footnote{
Notice the typo in the sign of the $\lambda_1^{SA}$ term
in the appendices of Refs.~\cite{L10,Natxo-thesis}.
} in the vector and axial-vector case:
\bear
\mF^v_{\varphi\varphi}(s) &=& 1\, +\, \sigma_V \;\Frac{s}{M_V^2 -s}\, .
\\
\mF^a_{S_1\varphi}(s) &=& \kw\; \left( \, 1\, +\, \sigma_A \:\Frac{s}{M_A^2 -s} \, \right)\, .
\eear

At high energy ($s\gg M_V^2, M_A^2, v^2$), the computed spectral functions behave as:
\begin{eqnarray}\label{eq:Pipp}
\left.\rho_S(s)\right|_{\varphi\varphi} & = &
\frac{g^2 \tan{\theta_W}}{192\varphi^2}\;\left\{\,
\left(1-\sigma_V\right)^2\, +\; {\cal O}(s^{-1})\,\right\}\, ,
\\[10pt]\label{eq:PiSp}
\left.\rho_S(s)\right|_{S_1\varphi} & = &\mbox{} -
\frac{g^2 \tan{\theta_W}}{192\pi^2}\;\kw^2\, \left\{\,
 \left(1-\sigma_A\right)^2\, +\; {\cal O}(s^{-1})\,\right\}\, .
\end{eqnarray}
Thus, their UV behaviour does not comply with the expected properties of the correlator $\Pi_{30}(s)$. At high energies, the total two-particle spectral function must behave as $\rho_S(s)\sim s^{-1}$. Furthermore, the first WSR would demand that this $s^{-1}$ term vanishes and the second WSR would require the $s^{-2}$ terms to be zero as well.

We will enforce that each of the two lowest-mass cuts, \ie the $\varphi\varphi$ and $S_1\varphi$ intermediate states, provides an acceptable representation of the (positive-definite) $\Pi_{VV}(s)$ and $\Pi_{AA}(s)$ correlators, respectively, at short distances. This means, that each of the two contributions should fall as $\cO(1/s)$, which implies that the form factors $\mF^v_{\varphi\varphi}(s)$ and
$\mF^a_{S_1\varphi}(s)$ should vanish at infinite momentum transfer. This condition determines the constraints:
\bear\label{eq.pipi-VFF-rel}
\sigma_V & \equiv & \frac{F_VG_V}{v^2} \;=\;1 \, ,
\\ \label{eq.Spi-AFF-rel}
\sigma_A & \equiv & \frac{F_A \lambda_1^{\mathrm{SA}}}{\kw v} \;=\; 1 \, ,
\eear
which imply a very smooth behaviour of $\rho_S(s)$.

Inserting the spectral function in the dispersion relation \eqn{eq.dispersivePi}, one obtains the real part of the correlator.
At short-distances, the resulting dispersive contribution behaves as
\be
\overline{\Pi}(s) \; =\;
\Frac{ v^2}{s} \; \delta_{_{\rm NLO}}^{(1)}\,
\, + \,
\Frac{v^2 \, M_V^2}{s^2}\;
\bigg[\delta_{_{\rm NLO}}^{(2)}
\, +\, \widetilde\delta_{_{\rm NLO}}^{(2)} \ln\Frac{-s}{M_V^2} \bigg]
\; +\;\cO\bigg(\Frac{1}{s^3}\bigg)\, ,
\label{eq.PI30-OPE}
\ee
where
\bear\label{eq:deltasNLO}
\delta^{(1)}_{\rm NLO}  &=&
 \Frac{M_V^2}{48\pi^2 v^2} \,\left\{ 1\, - \,  \kw^2\, \Frac{M_A^2}{M_V^2}\,
 \left[ 1 + \cO\left(\Frac{m_{S_1}^2}{M_A^2}\right)\right]\right\}\, ,
\nn\\
\delta^{(2)}_{\rm NLO}   &=&
 \Frac{M_V^2}{48\pi^2v^2}\, \left\{ 1\, - \,  \kw^2\, \Frac{M_A^4}{M_V^4}\,
 \left[ 1 + \ln\Frac{M_A^2}{M_V^2}
 + \cO\left(\Frac{m_{S_1}^2}{M_A^2}\right)\right]\right\}\, ,
\nn\\
\widetilde\delta^{(2)}_{\rm NLO}   &=&\mbox{}
-  \Frac{M_V^2}{48\pi^2 v^2} \,\left\{ 1\, - \, \kw^2\, \Frac{M_A^4}{M_V^4}\,
 \left[ 1 + \cO\left(\Frac{m_{S_1}^2}{M_A^2}\right)\right]\right\} \, .
\eear

Neglecting the small corrections of $\cO(m_{S_1}^2/M_A^2)$, the two-particle cut contribution to the parameter $S$ is found to be:
\bear\label{eq:Soverline}
\overline{S}   &=&
 \Frac{1}{12\pi} \, \left\{
  \left(\ln\Frac{M_V^2}{m_{H}^2}-\Frac{17}{6}\right)
 - \,\kw^2\, \left(\ln\Frac{M_A^2}{m_{S_1}^2}-\Frac{17}{6}\right)
\right\}   \, .
\eear

\subsection{First Weinberg sum rule at NLO}

The first Weinberg sum rule enforces the $\mO(1/s)$ term in $\widetilde{\Pi}_{30}(s)$ to vanish. Therefore, the term proportional to $\delta^{(1)}_{\rm NLO}$ in \eqn{eq.PI30-OPE} should cancel with the pole contributions in Eq.~\eqn{eq.T-NLO}. This gives the NLO relation~\cite{L8,L10,L8-Trnka,S-Higgsless}
\begin{eqnarray}
F_{V}^{r\,2}\, -\, F_{A}^{r\,2}\; =\; v^2\, \left(1\,+\,\delta_{_{\rm NLO}}^{(1)}\right)\, .
\label{eq:NLO_WSR1}
\end{eqnarray}

We have already seen at LO that imposing only the first WSR is not enough to determine the vector and axial-vector couplings. In that case, one can only
derive bounds on the $S$ parameter. Using the relation~\eqn{eq:NLO_WSR1} in Eq.~\eqn{eq.Sbar}, and assuming $M_A^r > M_V^r$, we obtain the inequality:
\begin{equation}
S \; =\; 4\pi \left\{
\frac{v^2}{M_V^{r\, 2}}\, \left( 1+\delta_{_{\rm NLO}}^{(1)}\right)
+ F_A^{r\,2} \left( \frac{1}{M_V^{r\, 2}} - \frac{1}{M_A^{r\, 2}} \right)
\right\} \, +\, \overline{S}
\; >\; \frac{4\pi v^2}{M_V^{r\, 2}}\,  \left( 1+\delta_{_{\rm NLO}}^{(1)}\right)
\, +\, \overline{S}
\, ,
\label{eq.NLO-S+1WSR}
\end{equation}
which at LO reduces to Eq.~\eqn{eq.LO-S+1WSR}. Substituting the one-loop results in Eqs.~\eqn{eq:deltasNLO} and \eqn{eq:Soverline}, one obtains:
\bear
S &> &  \Frac{4 \pi v^2}{M_{V}^{2}}
 \,\,+\,\,
\Frac{1}{12\pi}
\left[ \left(\ln\Frac{M_V^2}{m_{H}^2}-\Frac{11}{6}\right)
 - \,\kw^2\, \left(\log\Frac{M_A^2}{m_{S_1}^2}-\Frac{17}{6}
 + \Frac{M_A^2}{M_V^2}\right) \right]\, ,
\label{eq.S-NLO-1WSR}
\eear
where we have identified the LO and renormalized masses, \ie $M_{V,A}= M_{V,A}^r$, and terms of $\mathcal{O}(m_{S_1}^2/M_{A}^2)$ have been neglected.

Taking $m_H=m_{S_1}$, one finds
\bear
S &> &  \Frac{4 \pi v^2}{M_{V}^2}
 \,\,+\,\,\Frac{1}{12\pi}
\left[     \bigg(1-\kw^2 \bigg) \bigg(\log\Frac{M_V^2}{m_{S_1}^2}-\Frac{11}{6}\bigg)
 - \,\kw^2\, \bigg(\log \Frac{M_A^2}{M_{V}^2}-1
 + \Frac{M_A^2}{M_V^2}\bigg) \right]\, .
\label{eq.S-NLO-1WSRbis}
\eear
Thus, there are deviations from the LO lower bound when
either  $\kw\neq 1$ or $M_V\neq M_A$.

In the limit  $M_A\to M_V $, the inequality becomes an identity and one finds the simpler expression:
\bear
\lim_{M_A\to M_V} S & = &  \Frac{4 \pi v^2}{M_{V}^2}
 \,\,+\,\,
\Frac{1}{12\pi}\, \left(1-\kw^2\right)\,
\left[ \log{\Frac{M_V^2}{m_{S_1}^2}}-\Frac{11}{6} \right]\, .
\label{eq.S-1WSR-degenerate}
\eear

As it happened in the LO case, if we consider an inverted hierarchy of the vector and axial-vector resonances~\cite{Appelquist:1998xf,Marzocca:2012zn}, $M_A < M_V$, Eq.~\eqn{eq.NLO-S+1WSR} becomes an upper bound and all the inequalities flip direction:
\begin{equation}
S \; <\; \frac{4\pi v^2}{M_V^2}\;\left( 1+\delta_{\mathrm{NLO}}^{(1)}\right) + \overline{S} \, .
\label{eq.S-1WSR-inverted}
\end{equation}

\subsection{Second Weinberg sum rule at NLO}

The second Weinberg sum rule requires the stronger condition that $\widetilde{\Pi}_{30}(s)$ should fall as $\mO(1/s^3)$ at short distances. This is only possible if the
$\mO(\log{(-s)}/s^2)$ term in \eqn{eq.PI30-OPE} vanishes. Neglecting small corrections of $\cO(m_{S_1}^2/M_A^2)$, the constraint $\widetilde\delta^{(2)}_{\rm NLO} = 0$ relates the ratio of heavy resonance masses with the scalar coupling:
\be\label{eq:2WSR-a}
\kw = \Frac{M_V^2}{M_A^2}\, .
\ee
Since the LO WSRs have established the mass ordering $M_V<M_A$, the scalar coupling becomes bounded in the form $0< \kw <1$.
In addition, the cancellation of the $\mO(1/s^2)$ term in \eqn{eq.PI30-OPE} with the pole contributions at NLO implies the relation~\cite{L8,L10,L8-Trnka,S-Higgsless}
\begin{eqnarray}
F_{V}^{r\,2}\, M_{V}^{r\,2} \, -\, F_{A}^{r\,2}\, M_{A}^{r\,2} \; = \;
v^2 \, M_{V}^{r\,2} \,\delta_{_{\rm NLO}}^{(2)}  \, .
\label{eq:NLO_WSR2}
\end{eqnarray}

If one assumes the validity of the two WSRs it is then possible to fix the renormalized vector and axial-vector couplings in the form,
\begin{eqnarray}
F_{V}^{r\,2}&=& v^2\; \frac{M_{A}^{r\,2}}{M_{A}^{r\,2}-M_{V}^{r\,2}}\;
        \left(1+\delta_{_{\rm NLO}}^{(1)}-\frac{M_{V}^{r\,2}}{M_{A}^{r\,2}}\,\delta_{_{\rm NLO}}^{(2)}  \right) \, ,
\label{FVr} \nn\\[10pt]
F_{A}^{r\,2}&=& v^2\; \frac{M_{V}^{r\,2}}{M_{A}^{r\,2}-M_{V}^{r\,2}}\;
        \Bigl(1+\delta_{_{\rm NLO}}^{(1)}-\delta_{_{\rm NLO}}^{(2)} \Bigr) \, .
\label{FAr}
\end{eqnarray}
In the following,  we will use the renormalized masses $M_{R}^{r}$
in the NLO expressions and will denote them just as $M_R$.

Using Eqs.~\eqn{eq.Sbar} and \eqn{FAr}, one can fully determine the $S$ parameter in terms of the resonance masses:
\bear
S & =& \, 4\pi v^2 \, \bigg[\Frac{1}{M_{V}^{2}} +\Frac{1}{M_{A}^{2}}\bigg]
\;  \left(1+\delta_{\rm NLO}^{(1)}
-  \frac{M_{V}^{2}\, \delta_{\rm NLO}^{(2)}}{M_{V}^{2}+M_{A}^{2}} \right) \, \; +\,\;\overline{S}\, ,
\nonumber\\
&= &   4 \pi v^2 \left(\frac{1}{M_{V}^2}+\frac{1}{M_{A}^2}\right)
\, + \,  \frac{1}{12\pi}\,
\left[ \log\frac{M_V^2}{m_{H}^2}  -\frac{11}{6}
\right.\nonumber  \\ &&\left.
\qquad \qquad\qquad \qquad\qquad  \qquad\,
+\;\frac{M_V^2}{M_A^2}\log\frac{M_A^2}{M_V^2}
 - \frac{M_V^4}{M_A^4}\, \bigg(\log\frac{M_A^2}{m_{S_1}^2}-\frac{11}{6}\bigg) \right]  \,,
\label{eq.1+2WSR}
\eear
where terms of $\mathcal{O}(m_{S_1}^2/M_{V,A}^2)$ have been neglected and the relation \eqn{eq:2WSR-a} has been used.
The NLO spectral functions involve seven {\it a priori } unknown parameters:
$M_V$, $M_A$, $F_V$, $F_A$, $G_V$, $\kw$, and $\lambda_1^{\mathrm{SA}}$.
We have been able to determine five of them through the short-distance constraints in Eqs.~\eqn{eq.pipi-VFF-rel}, \eqn{eq.Spi-AFF-rel}, \eqn{eq:NLO_WSR1}, \eqn{eq:2WSR-a} and \eqn{eq:NLO_WSR2}. Therefore, only two free parameters remain in our final result.

Taking $m_H=m_{S_1}$, this expression can be further simplified to the form:
\be
S \; = \;   4 \pi v^2 \left(\frac{1}{M_{V}^2}+\frac{1}{M_{A}^2}\right)
\, + \,  \frac{1}{12\pi}\,
\left[ \left( 1-\frac{M_V^4}{M_A^4} \right) \left( \log\frac{M_V^2}{m_{S_1}^2}  -\frac{11}{6}\right)
+\;\left(\frac{M_V^2}{M_A^2}-\frac{M_V^4}{M_A^4} \right) \log\frac{M_A^2}{M_V^2} \right] .
\label{eq.1+2WSRbis}
\ee
The correction to the LO result vanishes when $M_V\to M_A$ ($\kw \to 1$). In this limit, the one-loop contributions cancel out and one recovers Eq.~\eqn{eq.LO-S+2WSR}.

\section{NLO calculation of $T$}
\label{sec:T-NLO}

\begin{figure}
\begin{center}
\includegraphics[scale=0.6]{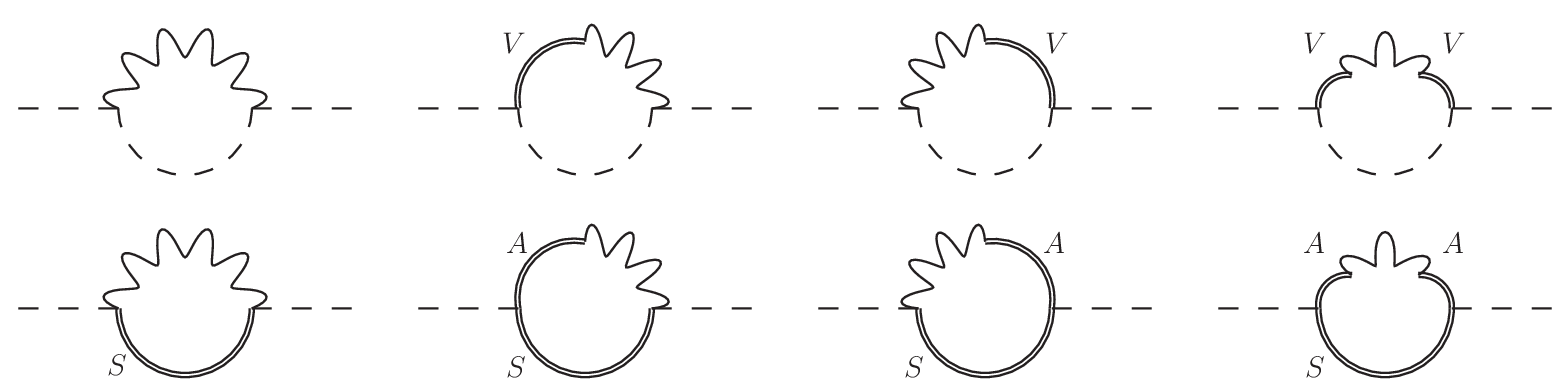}
\caption{\small{Absorptive diagrams contributing to the
Goldstone self-energies  and  the $T$ parameter at NLO.
A dashed (double) line stands for a Goldstone (resonance $V,A,S_1$) boson
and a curved line represents a $B$ gauge boson.
The first line provides the one-loop charged Goldstone self-energy
$-i\Sigma(s)^{(+)}$
and the second one the neutral one $-i\Sigma(s)^{(0)}$.
}}
\label{NLO_graphs}
\end{center}
\end{figure}

Figure~\ref{NLO_graphs} shows the computed one-loop contributions to $T$ from the lightest two-particle cuts. The self-energy of the charged Goldstone receives a non-zero contribution from loops with a Goldstone and a $B$ gauge boson, while the contributions to the neutral self-energy originate in a $S_1B$ cut.
The vertices required for the study of these cuts are the same we already used in the computation of $\Pi_{30}(t)$ for the $S$ parameter.

At the one-loop level both self-energies show a similar structure:
\bear
\left.\Sigma(q^2)^{(+)}\right|_{\varphi B} &=& g'^{\, 2}\, q_\mu q_\nu\;  \Int\; \Frac{{\rm d^Dk}}{i(2\pi)^D}
\; \left|\mF^{v}_{\varphi\varphi}(k^2)\right|^2 \;
\Frac{ g^{\mu\nu}-k^\mu k^\nu /k^2 }{k^2\, (q-k)^2}\, ,
\nn\\
\left.\Sigma(q^2)^{(0)}\right|_{S_1 B} &=&  g'^{\, 2}\, q_\mu q_\nu
\; \Int\; \Frac{{\rm d^Dk}}{i(2\pi)^D}
\, \left|\mF^{a}_{S_1\varphi}(k^2)\right|^2 \;
\Frac{ g^{\mu\nu}-k^\mu k^\nu /k^2 }{k^2  \,
  [(q-k)^2-m_{S_1}^2]     }
\, .
\label{eq.Sigma-Bpi}
\eear

Thus, the same $\varphi\varphi$ vector and $S_1\varphi$ axial-vector form factors entering the calculation of $S$ determine the one-loop contributions to $T$. Once the conditions \eqn{eq.pipi-VFF-rel} and \eqn{eq.Spi-AFF-rel} have been implemented,
the two form factors are very well behaved at high energies, implying also a good UV convergence of the Goldstone self-energies.\footnote{
This agrees with the observation made in Ref.~\cite{S-Orgogozo:11} that a well
behaved vector form factor at high energies led to a  cancellation of the UV divergences in their one-loop calculation of $T$.}
This allows us to perform an unambiguous determination of $T$ in terms of the resonance masses and $\kw$.

At low energies, the $\varphi B$ loop matches exactly the SM result; therefore, the dispersion relation~(\ref{eq.T-disp-rel}) is infrared safe and the $T$ parameter is well defined.
Notice as well that in the SM case the integral is UV finite due to the
cancellation between the $\varphi B$ and $S_1 B$   loops at short distances,
whereas in our strongly-coupled approach  each channel vanishes on its own at high energies~\cite{PRL}.
Neglecting terms of $\mathcal{O}(m_{S_1}^2/M_{A}^2)$, we obtain
\begin{equation}
 T\; =\;  \frac{3}{16\pi \cos^2 \theta_W}\; \left[ 1 + \log{\frac{m_{H}^2}{M_V^2}}
 - \kw^2\, \left( 1 + \log{\frac{m_{S_1}^2}{M_A^2}} \right)  \right]  \, ,
\label{eq:T}
\end{equation}
where $m_H$ is the SM reference Higgs mass adopted to define $S$ and $T$.
Taking $m_H=m_{S_1}$, one gets the simplified expression
\begin{equation}
 T\; =\;  \frac{3}{16\pi \cos^2 \theta_W}\; \left[ \left( 1-\kw^2 \right) \left( 1 + \log{\frac{m_{S_1}^2}{M_V^2}} \right)
 - \kw^2\, \log{\frac{M_V^2}{M_A^2}}   \right]  \, .
\label{eq:T-mS=mH}
\end{equation}
Therefore with $\kw=1$ (the SM value), $T$ vanishes when $M_V=M_A$ as it should.

\section{Constraints from electroweak precision data}
\label{sec.pheno}

\begin{figure}
\begin{center}
\includegraphics[scale=0.6]{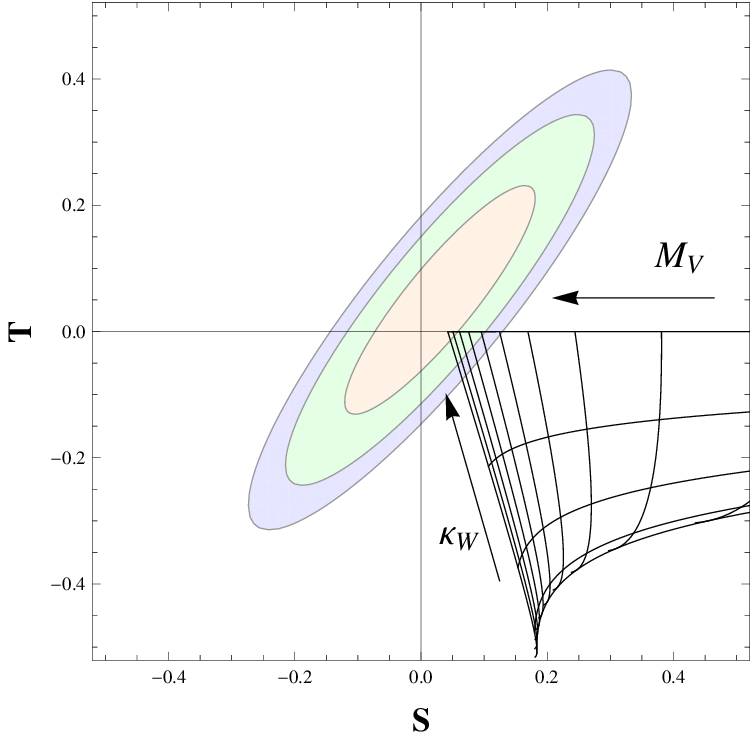}
\caption{\small{NLO determinations of $S$ and $T$, imposing the two WSRs.
The grid lines correspond to $M_V$ values from $1.5$ to $6.0$~TeV, at intervals of $0.5$~TeV, and $\kw= 0, 0.25, 0.50, 0.75, 1$.
The arrows indicate the directions of growing  $M_V$ and $\kw$.
The ellipses give the experimentally allowed regions at 68\% (orange), 95\% (green) and 99\% (blue) CL.}}
\label{fig.2WSR}
\end{center}
\end{figure}

In Figure~\ref{fig.2WSR}~\cite{PRL}
we show the compatibility between the experimental constraints on the parameters $S$
and $T$, given in Eq.~\eqn{eq:S_T_ewfit}, and our NLO determinations
in Eqs.~\eqn{eq.1+2WSRbis} and \eqn{eq:T-mS=mH}, with $\kw= M_V^2/M_A^2$,
imposing the two WSRs.
The line with $\kw=1$ ($T=0$) coincides with the LO upper bound in~(\ref{SLOtwoWSR}),
while the $\kw \,= M_V^2/M_A^2\to 0$ curve reproduces the lower bound
in Eq.~(\ref{eq.S-NLO-1WSRbis}) in the same limit.
Thus, a vanishing scalar-Goldstone coupling ($\kw=0$) would be incompatible with the data, independently of whether the second WSR has been assumed.

Figure~\ref{fig.2WSR} shows a very important result in the two-WSR scenario: with $m_{S_1} = 126$~GeV, the precision electroweak data requires that the Higgs-like scalar should have a $WW$ coupling very close to the SM one. At 68\% (95\%) CL, one gets
$\kw\in [0.97,1]$  ($[0.94,1]$),
in nice agreement with the present LHC evidence \cite{Aad:2012tfa,Aad:2013wqa,Chatrchyan:2012ufa,Chatrchyan:2013lba}, but much more restrictive. Moreover, the vector and axial-vector states should be very heavy (and quite degenerate); one finds $M_V> 5$~TeV ($4$~TeV) at 68\% (95\%) CL.

\begin{figure}
\begin{center}
\includegraphics[scale=0.75]{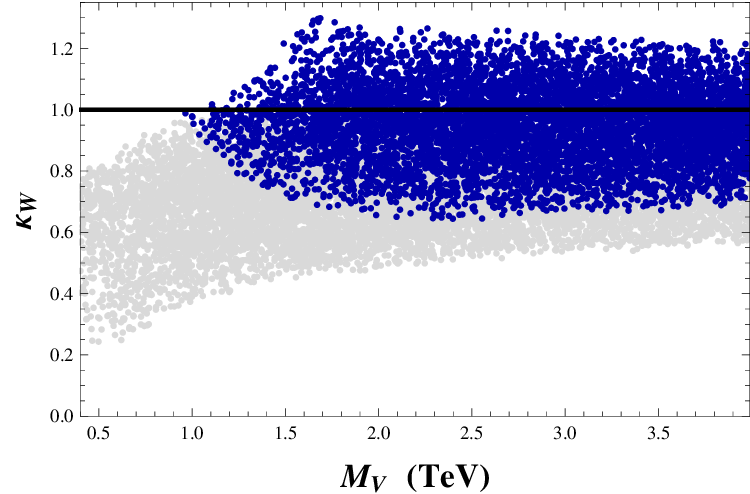}
\caption{\small Scatter plot for the 68\% CL region,
in the case when only the first WSR is assumed. The dark blue and light gray regions
correspond, respectively,  to $0.2<M_V/M_A<1$ and $0.02<M_V/M_A<0.2$.
We consider $M_A>M_V>0.4$~TeV in the plot.}
\label{fig.1WSR}
\end{center}
\end{figure}

This conclusion is softened when the second WSR is dropped and the lower bound in Eq.~\eqn{eq.S-NLO-1WSRbis} is used instead. This is shown in Figure~\ref{fig.1WSR}~\cite{PRL}, which gives the allowed 68\% CL region in the space of parameters $M_V$ and $\kw$, varying $M_V/M_A$ between 0 and 1. Note, however, that values of $\kw$ very different from the SM can only be obtained with a large splitting of the vector and axial-vector masses. In general there is no solution for $\kw >1.3$. Requiring $0.2<M_V/M_A<1$, leads to $1-\kw<0.4$
at 68\% CL,  while the allowed vector mass stays above 1~TeV~\cite{Filipuzzi:2012bv}.
Taking instead  $0.5<M_V/M_A<1$, one gets the stronger constraints
$1-\kw <0.16$ and $M_V>1.5$~TeV.
In order to allow vector masses below the TeV scale, one needs
a much larger resonance-mass splitting,
so that the NLO term in (\ref{eq.S-NLO-1WSRbis}) proportional to $\kw^2$
compensates the growing of the LO vector contribution.
The mass splitting gives also an additive contribution to $T$ of the form
$\delta T\sim \kw^2 \log{(M_A^2/M_V^2)}$,
making lower values of $\kw$ possible for smaller $M_V$.
However, the limit $\kw\to 0$ can only be approached
when $M_A/M_V\to \infty$.

\begin{figure}
\begin{center}
\includegraphics[scale=0.75]{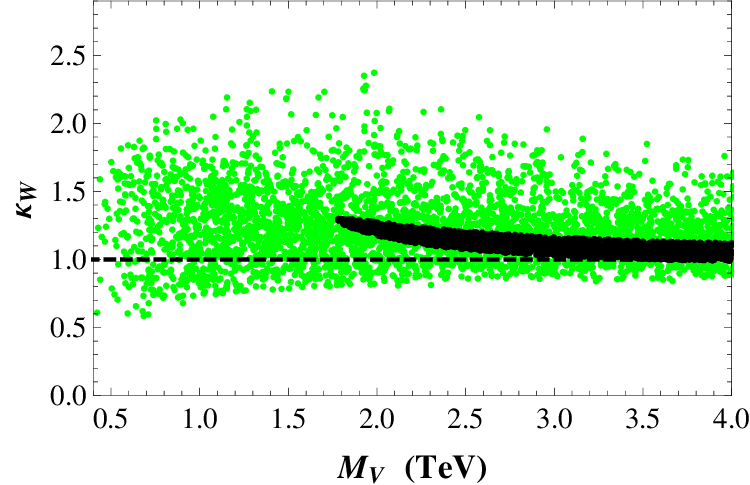}
\caption{\small
Scatter plot for the 68\% CL region,
in the case when only the first WSR is assumed, in the degenerate and inverted-hierarchy  scenarios.
The black (dark) and green (lighter) regions
correspond, respectively,  to
$M_V=M_A$ and $1<M_V/M_A<5$.
We consider $M_V>M_A>0.4$~TeV in the plot.}
\label{fig.1WSR-inverted}
\end{center}
\end{figure}

One may wonder what is the importance of assuming the normal hierarchy $M_V< M_A$,  as done in Figure~\ref{fig.1WSR}. Let us explore first the limit in which the spin--1 resonances are degenerate, $M_V=M_A$. The comparison with the experimental data yields the 68\% CL region plotted in black in Figure~\ref{fig.1WSR-inverted}. The allowed parameter space becomes very constrained around $\kw=1$, because both $S$ and $T$ put a limit on the difference $(1-\kw^2)$. One gets $0.97< \kw  < 1.30$, at the 68\% CL,
with $\kw$ getting closer to one for larger spin--1 resonance masses.
Moreover, the experimental constraints on the oblique parameters require  $M_V>1.8$~TeV at the 68\% CL.
The small width of the black band in Figure~\ref{fig.1WSR-inverted} is due to the experimental uncertainty on $S$ and $T$; it would shrink to a point if there were no errors. This can be easily checked, combining Eqs.~(\ref{eq.S-1WSR-degenerate}) and~(\ref{eq:T-mS=mH}) to eliminate the variable $\kw$. One gets then an implicit relation for $M_V$ in terms of
$S$ and $T$:
\be
M_V^2\; =\; 4\pi v^2\;\left\{ S + \Frac{4 \cos^2\theta_W}{9}\; T\;
\left(1-\frac{\frac{5}{6}}{\ln\frac{M_V^2}{m_H^2} -1}\right)\right\}^{-1}\, .
\label{eq.degenerate-Mv-1WSR}
\ee
For given $S$ and $T$, one gets a value of $M_V$ which inserted in~(\ref{eq:T-mS=mH}) determines $\kw^2$. Within present uncertainties, the denominator in \eqn{eq.degenerate-Mv-1WSR} is compatible with zero and, therefore, $M_V$ could take arbitrary large values.

The green region in Figure~\ref{fig.1WSR-inverted} shows the 68\% CL allowed area
in the inverted-hierarchy scenario with $M_A<M_V$. It continues the upper part
of the results for $M_V<M_A$ in Figure~\ref{fig.1WSR}, up to a slight overlap
due to the experimental errors. Now there are allowed solutions  extending below  $M_V=1$~TeV and beyond $\kw=1.3$. For a moderate splitting $1<M_V/M_A<5$  \  ($1<M_V/M_A<2$), the scalar coupling $\kw$ is nonetheless  constrained to the range $\kw<2.4$ \ ($\kw<2.0$). However, in the inverted hierarchy case there
can be spin--1 resonances with $M_A<M_V<1$~TeV
and $\kw$ can run down to zero if the masses are as small as $M_V\sim 0.5$~TeV. This corner of the parameter space ($M_{V,A}\simeq 0.5$~TeV, $\kw\simeq 0$), although possible, is nevertheless extremely disfavoured and can only be observed
with a much larger number of points in the scatter plot in Figure~\ref{fig.1WSR-inverted}.
On the other hand, if $M_V/M_A>5$ the cloud of (lighter green)
points in Figure~\ref{fig.1WSR-inverted}
gets shifted towards larger vector masses and higher scalar couplings,
with $M_V>2$~TeV and $\kw >1.1$ at the 68\% CL. A wider splitting,
with larger $M_V/M_A$, increases this allowed $\kw$ range even further.
Nevertheless, in the case of a moderate splitting
it is remarkable that if no vector resonance is observed below the TeV ($M_V>1$~TeV)
then the scalar coupling  becomes constrained  to values around $\kw \sim 1$,
as we found in the normal hierarchy case (Figure~\ref{fig.1WSR}).
More precisely, for $1<M_V/M_A<2$ we find  $0.7< \kw < 1.9$
at the 68\% CL  if the vector mass is over 1~TeV.

\section{Summary}
\label{sec.conclusions}

We have performed the first combined analysis of the oblique parameters $S$ and $T$,
including the impact of the newly discovered Higgs-like boson,
within an effective field theory framework including spin--1 resonances,
at the one-loop level.
We consider a general Lagrangian implementing the $SU(2)_L\otimes SU(2)_R\to SU(2)_{L+R}$ pattern of EWSB, with a non-linear realization of the corresponding Goldstone bosons.
The Lagrangian contains the lowest multiplets of vector and axial-vector resonance states, generated by the strongly-coupled underlying dynamics, and the Higgs-like boson with $m_{S_1}=126$~GeV is incorporated as a scalar singlet, without any further
specification about its origin. In this article we have completed the results presented in our previous letter~\cite{PRL}, and have given a detailed description of the adopted methodology.

Our results do not depend on unphysical UV cut-offs, widely used in previous literature~\cite{S-Isidori:08,S-Cata:10,S-Orgogozo:11,Orgogozo:2012}.
This is achieved trough the use of dispersion relations and proper short-distance constraints, reflecting the assumed UV properties of the underlying
strongly-coupled electroweak theory. Imposing a good high-energy behaviour for the $\Pi_{30}(s)$ correlator, one obtains finite dispersive integrals which can be calculated in terms of a few resonance parameters. We distinguish two different scenarios for the asymptotic fall-off at large momenta: the one obeyed by asymptotically free theories, which have very good UV properties (two WSRs; technicolour-like), and another one with a much weaker requirement (only the first WSR), expected to be satisfied in more general frameworks.

The light Higgs-like boson plays a very important role to compensate potentially large contributions from Higgsless channels (specially in the $T$ parameter). This effect is crucial to reproduce the electroweak precision observables, requiring in general a scalar coupling close to the SM one, \ie  $\kw\simeq 1$, and masses over the TeV scale. In the more restrictive scenario, where the two WSRs remain valid, we find at 68\% (95\%) CL:
\be
0.97 \; (0.94)\; <\; \kw\; <\; 1\, ,
\qquad\qquad
M_A\; >\; M_V\; >\; 5\; (4)\: \mathrm{TeV}.
\ee

These strong bounds get softened when only the first WSR is required to be valid.
On general grounds, one would expect this scenario to satisfy the mass hierarchy $M_V<M_A$. Assuming a moderate splitting $0.5<M_V/M_A<1$, we obtain (68\% CL)
\be
0.84 \; <\;\kw\; <\; 1.3\, ,
\qquad\qquad
M_V\; >\; 1.5\: \mathrm{TeV}.
\ee
Slightly larger departures from the SM can be achieved by considering a larger mass splitting. On the contrary, when the resonance masses become degenerate, the allowed range for $\kw$ reduces to $0.97 < \kw < 1.3$, and a heavier resonance mass is necessary, $M_V=M_A > 1.8$~TeV (68\% CL).

We have also analyzed the unlikely inverted-mass scenario, $M_V>M_A$,
finding that a large mass splitting is disfavoured by the  LHC data
on $\kw$~\cite{Aad:2013wqa,Chatrchyan:2013lba}.
For a moderate splitting $1<M_V/M_A<2$, we obtain  the upper bound
$\kw<2$ (68\% CL), while $\kw$ is allowed (though extremely disfavoured)  to be very small if the spin--1 masses are close to 0.5~TeV.
However, as soon as $M_V>1$~TeV, the scalar coupling  becomes  lower bounded:
$0.7 < \kw < 1.9$ (68\% CL).

All these results point out that, contrary to what is sometimes stated,
the current electroweak precision data easily allow for massive resonance states
at the natural electroweak scale, \ie well over the TeV. However, the scalar coupling $\kw$ is strongly constrained, specially for large resonance masses.
As the effect from heavy states becomes smaller, the light scalar is forced to couple to the gauge bosons with a value closer to the one of the SM Higgs coupling, in order to satisfy the experimental limits on $S$ and $T$.

Our conclusions are quite generic, since we have only used mild assumptions about the UV behavior of the underlying strongly-coupled theory, and can be easily particularized to more specific models obeying the
$SU(2)_L\otimes SU(2)_R\to SU(2)_{L+R}$ pattern of EWSB. An  example
is provided by the $SO(5)/SO(4)$ minimal composite
Higgs model~\cite{Orgogozo:2012,composite,Contino:2010rs,Marzocca:2012zn,Contino:2011},
where the scalar coupling is related to  the $SO(4)$ vacuum angle $\theta$
and upper bounded in the form $\kw=\cos\theta\leq 1$~\cite{composite,Contino:2010rs}.
With this identification, the $S$ and $T$ constraints in Figure~\ref{fig.2WSR}
remain valid in this composite scenario (see appendix~\ref{app.SO5-SO4}
for further details).
Another possibility would be to interpret the Higgs-like scalar as a dilaton,
the pseudo-Goldstone boson associated with the spontaneous breaking of scale
(conformal) invariance at a scale
$f_\phi \gg  v$~\cite{Goldberger:2007zk,Matsuzaki:2012xx,Bellazzini:2012vz,Chacko:2012vm,EP:2012}.
The dilaton coupling to the SM electroweak
bosons corresponds to $\kappa_W = v/f_\phi$, which makes this scenario with a high
conformal symmetry-breaking scale  quite unlikely.
The (fine-tuned) requirements needed to accommodate
a light dilaton with $\kappa_W \sim 1$ have been recently discussed in Ref.~\cite{Bellazzini:2012vz}.

\section*{Acknowledgments}

JJSC would like to thank A. Dobado, M.J. Herrero, A. Pomarol and J. Shu for useful discussions on chiral Lagrangians and composite Higgs models.
This work has been supported in part by the Spanish Government
and ERDF funds from the European Commission [grants 
FPA2010-17747, FPA2011-23778, AIC-D-2011-0818, SEV-2012-0249 (Severo Ochoa Program), CSD2007-00042 (Consolider Project CPAN)], the Generalitat Valenciana
[PrometeoII/2013/007] and the Comunidad de Madrid [HEPHACOS
S2009/ESP-1473].

\appendix
%

\section{Field redefinitions and higher-derivative operators}
\label{app.RChT-EoM}

At the one-loop level, the study of the absorptive contributions to $\Pi_{30}$
and the Goldstone self-energies only requires a limited amount of vertices with at most three-legs. Moreover, if we focus on the lightest absorptive channels to
$\rho_S(t)$ and $\rho_T(t)$, $\{ \, \varphi\varphi\, ,\, S_1\varphi\, \}$
and $\{\, B\varphi\, ,\, BS_1\, \}$, respectively,
only 6 kinds of local interactions are needed:
the two-leg transition vertices \ $W_3 , B \to  V$ \ and \ $W_3 , B \to  A$,
and the three-leg vertices \ $W_3 , B \to \varphi\varphi$, \
$W_3 , B \to S_1\varphi$, \ $V \to\varphi\varphi$ \ and \ $A \to S_1\varphi$.

Let us consider the most general effective Lagrangian contributing to these types of vertices, consistent with the assumed $SU(2)_L\otimes SU(2)_R\to SU(2)_{L+R}$ invariance.
We will not impose any further constraint on the allowed structures; thus, the Lagrangian could include operators with an arbitrary number of derivatives. Nevertheless,
the equations of motion (EoM) and the invariance of the generating functional under
field redefinitions can be used to reduce the number of relevant operators \cite{RChT}.
Following the strategy developed in Ref.~\cite{RChT-EoM}, we will consider appropriate
field redefinitions to reduce the number of derivatives on vertices of the needed type, up to structures with a higher number of fields which cannot contribute to our calculations.
The procedure consists on simplifying first the operators with two particle fields (up to remainders with three or more fields); then terms in the Lagrangian with three particle fields (up to remainders with four or more fields); and so on.
We will denote as $X$ to any resonance, Goldstone or gauge field
and $J$ will refer in general to one gauge boson field.

The starting point are the LO kinetic Lagrangians that provide the free canonical
propagators ($R=V,A$):
\bear
\mL^{\rm kin}_{\varphi\varphi} &=& \Frac{v^2}{4}\, \bra u_\mu u^\mu \ket \, ,\nn\\
\mL^{\rm  kin}_{S_1S_1} &=&   \Frac{1}{2} [\, \partial_\mu S_1 \, \partial^\mu S_1\,
-\, m_{S_1}^2  \, S_1^2\, ]\, ,
\nn\\
\mL^{\rm kin}_{RR} &=& \, -\, \Frac{1}{2}
\bra \nabla^\mu R_{\mu\rho} \nabla_\nu R^{\nu\rho}\ket
\, +\, \Frac{M_R^2}{4}\bra R_{\mu\nu} R^{\mu\nu}\ket \, .
\eear
These $SU(2)_L\otimes SU(2)_R$ invariant structures contain, in addition,
interaction terms.

The EoM are given by the variation of the whole action under infinitesimal
field redefinitions of the form
$S_1\to S_1 +\eta_{S_1}$, $R^{\mu\nu}\to R^{\mu\nu}+ \eta^{\mu\nu}_R$
and $u(\varphi)\to u(\varphi) \, \exp\{- i \eta_\varphi/4\}$.
At linear order in the variation one has
\bear
\bra \eta_\varphi\, \Frac{\delta \mS}{\delta\eta_\varphi} \ket  \;\, &=&\;\,
\Frac{v^2}{4}\, \bra \eta_\varphi\, \left[\nabla^\mu u_\mu \,+\,  \cO(X^2)\right]
\ket\, ,
\nn\\
\bra \eta_{S_1}\,
\Frac{\delta \mS}{\delta\eta_{S_1}}\ket   \;\, &=&\;\,
\, -\,  \eta_{S_1}\, \left[ \,
(\partial^2\,  +\,  m_{S_1}^2 )\, S_1 \,+\, \cO(X^2)\right] \, ,
\nn\\
\bra \eta_R^{\mu\nu}\Frac{\delta \mS}{\delta\eta_{R}^{\mu\nu}}\ket   \;\, &=&\;\,
\Frac{1}{2} \bra \eta_{R,\, \mu\nu}\, \left[
 \nabla^\mu\nabla_\rho R^{\rho\nu} \,- \, \nabla^\nu \nabla_\rho R^{\rho\mu}
\, +\, M_R^2 \, R^{\mu\nu}
\, +\, \cO(J)\, +\,\ \cO(X^2)\right]\ket \, .
\nn\\
\label{eq.S-var1}
\eear
Furthermore, if the spin--1 transformation $\eta_R^{\mu\nu}$
is chosen to be of the form
\bear
\eta_R^{\mu\nu} &=& \Frac{1}{2 M_R^2}
\, \left[\, (\nabla^2 +M_R^2) g_\alpha^\mu g_\beta^\nu
-\nabla_\alpha \nabla^\mu g_\beta^\nu
+\nabla_\beta \nabla^\mu g_\alpha^\nu
\, -\, (\mu\leftrightarrow \nu)\, \right]\, \hat{\eta}^{\alpha\beta}_R\, ,
\eear
then  at linear order in $\hat{\eta}_R^{\alpha\beta}$
one obtains the action variation
\bear
\bra \eta_R^{\mu\nu} \Frac{\delta \mS}{\delta\eta_{R}^{ \mu\nu}} \ket \;\, &=&\;\,
\Frac{1}{2}\,\bra \hat{\eta}_{R,\, \alpha\beta }\,
\left[  \, ( \nabla^2 + M_R^2)   R^{\alpha\beta}
\, +\, \cO(J)\, +\, \cO(X^2)\right]\ket \, .
\label{eq.S-var2}
\eear
We have used the property that the commutation of covariant derivatives
adds more fields: $[\nabla^\mu,\nabla^\nu]=\cO(J)+\cO(X^2)$.
Eqs.~\eqn{eq.S-var1} and~\eqn{eq.S-var2}  are identically zero when the fields 
are solutions of the classical EoM. 

Taking a convenient choice of finite field redefinitions $\eta_i$, these identities allow us to trade operators of the form \
$\bra \eta_\varphi \nabla^\mu u_\mu \ket$, \
$\eta_{S_1} \partial^2  S_1 $, \
$\bra \eta_{R,\, \mu\nu}
(\nabla^\mu\nabla_\rho R^{\rho\nu} - \nabla^\nu \nabla_\rho R^{\rho\mu})\ket$ \ and \
$\bra \hat{\eta}_{R}^{\alpha\beta} \nabla^2  R_{\alpha\beta} \ket$,
by operators with either the same number of fields and less derivatives,
operators with a higher number of fields or operators
where one spin--1 resonance field $R$ is replaced by a gauge boson $J$.

Let us analyze first the $J\to \varphi ,\, V,\, A$ transitions, with $J=W,\, B$.
Following the previous indications it is possible to simplify the contributing
action into the minimal basis~\footnote{
Contrary to what happens in the pure Goldstone theory, the
$W,B\to \varphi$ coupling might suffer a renormalization at NLO~\cite{RPP:05,L8-Trnka}.
We assume a renormalization scheme such that the renormalized coupling $v^r$ coincides with $v= 246$~GeV~\cite{Appelquist:1980vg,Longhitano:1980iz,Dobado:1990zh,Espriu:1991vm,Dobado:1997jx}.
}
\bear
\mL_{J\varphi}+\mL_{JV}+\mL_{JA} &=&
\Frac{v^2}{4}\, \bra u^\mu u_\mu \ket \, +\,
\Frac{F_V}{2\sqrt{2}}\, \bra V_{\mu\nu} f_+^{\mu\nu}\ket
\, +\, \Frac{F_A}{2\sqrt{2}}\, \bra A_{\mu\nu} f_-^{\mu\nu}\ket \, ,
\eear
generating a remainder of  operators with two gauge bosons
(which do not contribute to our problem at hand)
or three particle fields (which will be simplified next).

The second step is the analysis of the terms that participate in the transitions
$V\to \varphi\varphi$ and $A\to S_1\varphi$, without gauge bosons.
Through appropriate field redefinitions one may arrange the minimal basis
\bear
\mL_{V\varphi\varphi}+\mL_{AS_1\varphi} &=&
\Frac{i G_V}{2\sqrt{2}} \,\bra V_{\mu\nu} [u^\mu,u^\nu]\ket
\,+\, \sqrt{2} \lambda_1^{SA} \, \partial_\mu S_1\, \bra A^{\mu\nu} u_\nu\ket \, ,
\eear
at the price of generating terms with gauge bosons of $\cO(J X^2)$
and operators with four particle fields.

The third and final step is the obtention of a minimal basis of operators
for the $J\to \varphi\varphi$ and $J\to S_1\varphi$ transitions, but paying attention
of not spoiling the previous simplifications.
The analysis of all possible combinations of covariant tensors yields
the reduced Lagrangians
\bear
\mL_{J\varphi^2} &=& \Frac{v^2}{4}\,\bra u_\mu u^\mu\ket
\,\,\, +\,\,\,
\sum_{n\geq 0} i\, \lambda_n^{J\varphi\varphi}\,\,
\bra [u_\mu, u_\nu] \, (\nabla^2)^n f_+^{\mu\nu}\ket\, ,
\nn\\
\mL_{JS_1\varphi} &=& \Frac{\kw\, v}{2}\, S_1\, \bra u_\mu u^\mu\ket
\,\,\, +\,\,\,
\sum_{n\geq 0}  \lambda_n^{JS_1\varphi}\,\,
\partial_\mu S_1\, \bra  u_\nu  \, (\nabla^2)^n f_-^{\mu\nu}\ket\, ,
\eear
generating a remainder of operators with two gauge fields
or terms with four or more particle fields.

We could not find a way to further reduce the operator basis through
field redefinitions. However, the study of the $\varphi\varphi$ vector and
$S_1\varphi$ axial form factors yields at LO
\bear
\mF^v_{\varphi\varphi}(s) &=& \, 1\, +\, \sigma_V \, \Frac{s}{M_A^2-s}
\,\,\,+ \,\,\, \sum_{n\geq 0} \Frac{4\lambda_n^{J\varphi\varphi}}{v^2}\, (-s)^{n+1}  \, ,
\nn\\
\mF^a_{S_1\varphi}(s) &=&
\kw\, \bigg[ \, 1\, +\, \sigma_A \, \Frac{s}{M_A^2-s}
\,\,\,+\,\,\, \sum_{n\geq 0} \Frac{\lambda_n^{JS_1\varphi}}{\kw v}\, (-s)^{n+1}
\,\bigg]\, ,
\eear
with $\sigma_V=F_V G_V/v^2$ and $\sigma_A= F_A\lambda_1^{SA}/(\kw v)$.
The requirement that these two form-factors vanish at high momentum
leads to the resonance constraints previously quoted in~(\ref{eq.pipi-VFF-rel})
and~(\ref{eq.Spi-AFF-rel}), together with the absence of higher
derivative operators of the form $J\varphi\varphi$ and $JS_1\varphi$:
\bear
&  \sigma_V\, =\,  \sigma_A\, =\, 1 \, , \qquad \qquad\qquad
\lambda_n^{J\varphi\varphi}\,=\, \lambda_n^{JS_1\varphi}\,=\, 0\, .
\eear

\section{$\mathbf{SO(5)/SO(4)}$ composite models}
\label{app.SO5-SO4}

This kind of models assumes the spontaneous symmetry breaking $SO(5)\to SO(4)$,
at some high-energy scale $4\pi f$, which results in the appearance of four Goldstone bosons, one for each broken generator. Three of them correspond to the usual electroweak Goldstones, while the fourth one is identified with the light Higgs-like boson.
In order to account for the Higgs mass, $m_{S_1}\simeq 126$~GeV, one further assumes that the vacuum becomes misaligned through some dynamical mechanism
(e.g. radiative corrections in extra dimensions~\cite{composite,Contino:2011}), so that the fourth Goldstone gains a small mass, much smaller than the EWSB scale $4\pi v$,
and becomes a pseudo-Goldstone field. The vacuum misalignment is determined by the
ratio of the two symmetry-breaking scales:  $v/f=\sin \theta \leq 1$.

The four Goldstones are non-linearly realized and the action is constructed
by means of the standard CCWZ formalism~\cite{Coleman:1969sm}.
The $\cO(p^2)$ Goldstone Lagrangian is given
by~\cite{composite,Contino:2011,Marzocca:2012zn}
\be
\mL\; =\; \Frac{f^2}{4}\, \sin^2(\theta +h(x)/f)\;\bra u_\mu u^\mu\ket
\; =\; \Frac{v^2}{4}\,
\left(1 + \frac{2}{v}\, h(x) \cos\theta  \right)\; \bra u_\mu u^\mu\ket +\cO(h^2)\, ,
\ee
which has exactly the same structure as our electroweak effective Lagrangian in Eqs.~(\ref{eq.L_G}) and \eqn{eq:L_R}, 
with the $S_1\varphi\varphi$ interaction given by
\be
\kw\,=\, \cos\theta\, .
\ee

The interaction between a spin--1 $SO(5)$ resonance $\rho$ and the $SO(5)/SO(4)$ Goldstones takes the generic form~\cite{Contino:2011}   
\bear\label{eq:R_SO5}
\mL &=& \Frac{K_F}{2\sqrt{2}} \,\bra \rho_{\mu\nu} f_+^{\mu \nu}\ket \, +\,
 \Frac{i K_G}{2\sqrt{2}} \,\bra \rho^{\mu\nu} [ d_\mu, d_\nu]\ket  \, ,
\eear
where the chiral tensors here refer to $SO(5)/SO(4)$~\cite{Contino:2011}.
The $SO(5)$ $\rho$ multiplet contains both vector and axial-vectors states:
$\rho^{\mu\nu}= \rho^{\mu\nu,a}_V T^{a}_V + \rho^{\mu\nu,a}_A T^{a}_A$,
with
$T^{a}_V=(T^{a}_R+T^{a}_L)/\sqrt{2}$ and $T^{a}_A=(T^{a}_R-T^{a}_L)/\sqrt{2}$.
The structure \eqn{eq:R_SO5} reproduces the electroweak Lagrangian in Eqs.~\eqn{eq:L_R}, with the identifications:
\bear\label{eq:SO5relations}
F_V\!\! & = &\!\! K_F\, , \hskip 3.5cm   G_V\; =\; \Frac{1}{2}\, K_G\, \sin^2\theta\, ,
\nonumber\\
F_A &\!\! = &\!\! K_F\,\cos\theta \,  , \hskip 2.5cm   \lambda_1^{SA} v \; =\;
\Frac{1}{2}\, K_G\, \sin^2\theta\, .
\eear

\subsection{High-energy constraints}

A fully symmetric $\rho$ multiplet containing vector and axial-vector states fulfills the short-distance conditions in a very natural way. Since the couplings $K_F$ and $K_G$ are common to the whole multiplet, the vector and axial-vector form-factor constraints in Eqs.~\eqn{eq.pipi-VFF-rel} and \eqn{eq.Spi-AFF-rel} generate the same relation:
\be\label{eq:V-A-FF}
K_F K_G \,=\, 2 f^2\, .
\ee
When this condition is satisfied, the two form factors follow automatically the same
high-energy behaviour.

At LO the first WSR in Eq.~\eqn{eq:1stWSR-LO} implies $K_F=f$. Together with Eq.~\eqn{eq:V-A-FF}, this implies $K_G = 2f$. The relations \eqn{eq:SO5relations} determine then the couplings of the $SO(5)/SO(4)$ Lagrangian to take the values
$F_V=f$, \ $F_A=f\, \cos{\theta}$, \
$G_V = f \sin^2{\theta}$ \ and \ $\lambda_1^{SA} = \sin{\theta}$.

The second WSR in \eqn{eq:2ndWSR-LO} requires $\cos^2{\theta}= M_V^2/M_A^2$.
A symmetric $\rho$ multiplet with $M_V = M_A$ would imply $\cos^2{\theta}=1$ and, therefore, $v/f =\sin{\theta}=0$. We must then allow for a small splitting of $\cO(\sin^2\theta)$ in the multiplet $\rho=(\rho_V,\rho_A)$. In fact, although the second WSR predicts $M_V\le M_A$, as expected, the resulting LO condition differs from the NLO constraint in Eq.~\eqn{eq:2WSR-a} which implies $\cos{\theta} = M_V^2/M_A^2$.
The difference between both expressions is indeed of $\cO(\sin^2\theta)$, and can easily
be accounted for through a small splitting of that order in the vector and axial masses and couplings.  
Let us parametrize the splitting in the form:
\be\label{eq.FR+splitting}
M_A^2\; =\; M_V^2\, \left(1+\epsilon_M\right)\, ,
\qquad\quad
F_V^2\; =\; K_{F,0}^2 \, \left(1+\epsilon_V\right)\, ,
\qquad\quad
F_A^2\; =\; K_{F,0}^2\, \cos^2{\theta} \, \left(1+\epsilon_A\right) \, ,
\ee
with $\epsilon_i\to 0$ ($i=M,V,A$) for  $\theta\to 0$
(and $F_{V,A}\to K_{F,0}$). Then the first WSR allows a general value
for $K_{F,0}$ and constrains the splitting in the form
\begin{equation}
(\epsilon_V-\epsilon_A) \; =\; \left(\Frac{f^2}{K_{F,0}^2} - 1\right)  \sin^2{\theta}
-  \epsilon_A\sin^2{\theta} \;
= \;  \left(\Frac{f^2}{K_{F,0}^2} - 1\right)  \sin^2{\theta}
+\cO(\theta^3)\, .
\label{eq.splitting-epsV-epsA}
\end{equation}
The application of the first and second WSRs leads to a prediction of the resonance couplings in terms of the resonance masses:
\bear
\Frac{F_V^2}{v^2}\; =\; \Frac{M_A^2}{M_A^2-M_V^2}\; =\; \Frac{1}{\epsilon_M} +1 \, ,
\qquad\qquad
\Frac{F_A^2}{v^2}\; =\; \Frac{M_V^2}{M_A^2-M_V^2}\; =\; \Frac{1}{\epsilon_M} \, .
\label{eq.composite-1+2WSR+splitting}
\eear
By means of  the values of $F_V$ and $F_A$ in Eq.~(\ref{eq.FR+splitting}),
one extracts the LO determination,
\bear
\epsilon_M\; =\; \Frac{f^2}{K_{F,0}^2}\sin^2\theta \,+\,\cO(\theta^3)\, .
\eear

This kind of models are particularly interesting in the present phenomenological situation,
where LHC has found a  Higgs-like boson much lighter than the electroweak scale
$\Lambda_{EW}=4\pi v\sim 3$~TeV
and nothing else so far. This could be an indication that this scalar
is a pseudo-Goldstone boson of some
global symmetry beyond the SM, among which, $SO(5)$ is
the simplest extension which embeds the $SO(4)\sim SU(2)_L\otimes SU(2)_R$ group
and may have four generators spontaneously broken.
The assumption of this symmetry pattern naturally reproduces  the most favoured results
of the $S$ and $T$ phenomenological analysis in the present article:
small $V-A$ splitting and $\kw\sim 1$; important cancelations between 
Higgsless (Goldstone) channels and cuts with scalars, as all the four constitute a full multiplet of Goldstone bosons (or pseudo-Goldstones, in the $S_1$ case); 
and a well-defined perturbative framework with small loop corrections up to energies beyond
$\Lambda_{EW}$, suggesting the presence of a higher scale $4\pi f$ suppressing the loops.
Nonetheless, we remind the reader that
all along the work  we worked within a general framework, leaving the couplings unfixed, and the relations in this appendix were not considered.


\end{document}